\documentclass[11pt,a4paper]{article}
\pdfoutput=1
\usepackage{jheppub}

\usepackage{float}
\usepackage{amsmath}
\usepackage{cleveref}
\usepackage{xcolor}

\newcommand{\ud}{\mathrm{d}}

\crefname{equation}{eqn.}{eqns.}
\Crefname{equation}{Eqn.}{Eqns.}
\crefname{figure}{fig.}{figs.}
\Crefname{figure}{Fig.}{Figs.}

\usepackage{mathdots}
\usepackage{arydshln}

\title{ Real-time quantum dynamics, path integrals and the method of thimbles.
}

\author[b]{Zong-Gang Mou,}
\author[a]{Paul M. Saffin,}
\author[b]{Anders Tranberg,}
\author[a]{Simon Woodward}

\affiliation[a]{School of Physics and Astronomy, University Park, University of Nottingham,\\ Nottingham NG7 2RD, United Kingdom}
\affiliation[b]{Faculty of Science and Technology, University of Stavanger, 4036 Stavanger, Norway}

\emailAdd{zonggang.mou@uis.no}
\emailAdd{paul.saffin@nottingham.ac.uk}
\emailAdd{anders.tranberg@uis.no}
\emailAdd{simon.woodward1@nottingham.ac.uk}

\abstract{ Direct numerical evaluation of the real-time path integral has a well-known sign problem that makes convergence exponentially slow. One promising remedy is to use Picard-Lefschetz theory to flow the domain of the field variables into the complex plane, where the integral is better behaved. By Cauchy's theorem, the final value of the path integral is unchanged. Previous analyses have considered the case of real scalar fields in thermal equilibrium, employing a closed Schwinger-Keldysh time contour, allowing the evaluation of the full quantum correlation functions. Here we extend the analysis by not requiring a closed time path, instead allowing for an initial density matrix for out-of-equilibrium initial value problems. We are able to explicitly implement Gaussian initial conditions, and by separating the initial time and the later times into a two-step Monte-Carlo sampling, we are able to avoid the phenomenon of multiple thimbles. In fact, there exists one and only one thimble for each sample member of the initial density matrix. We demonstrate the approach through explicitly computing the real-time propagator for an interacting scalar in 0+1 dimensions, and find very good convergence allowing for comparison with perturbation theory and the classical-statistical approximation to real-time dynamics. }

\begin{document}

\maketitle

\section{Introduction}
\label{sec:Intro}

{For a quantum system evolving from one state to another, Feynman's path integral quantization asserts that all possible paths in field space contribute to the quantum amplitude. These contributions are equal in magnitudes but have different phases \cite{feynman}.
This poses a great challenge when one wants to compute the path integral through numerical methods, as although the interesting physics is often concentrated in some region of space of paths, a detailed cancellation of quickly oscillating functions must be achieved. The challenge is also known as the ``sign problem'', and appears whenever the path integral kernel cannot be made real by Wick rotation to a Euclidean action, such as when a chemical potential is present, or the correlators one is trying to compute involve a real time separation.} 

Recently in \cite{Alexandru:2016gsd,Alexandru:2017lqr}, it was shown that the real-time path integral can be computed through a Generalized Thimble Method, based on complexifying the field variables. In the complexified field space, one can deform the integration cycle of the path integral into the complex plane and still obtain the same result of the integral, so long as the integrand is holomorphic in the new complex variables. There is some freedom in how one deforms the contours, but a natural choice is to use a gradient flow (to be described below), starting from the original real field space. If we stop the flow at some finite flow-time, then the original field space will have flowed to some new field space ${\cal M}$, and the integral over ${\cal M}$  is equivalent to the integral over original real field space.
In particular, when the flow time approaches infinity, one finds that ${\cal M}_\infty$ is composed of Lefschetz thimbles \cite{Witten:2010cx}, where the phase of the integrand is constant and therefore the ``sign problem'' is eliminated on each thimble (except for milder contributions from the residual phase  \cite{Cristoforetti:2012su}).
In practise, we are not able to perform infinitely long flows. However, as long as the flow-time is large enough then the ``sign problem'' will be alleviated, in the sense that the highly oscillatory integrals of a function with constant magnitude (i.e. $e^{i S/\hbar}$) turn into integrals of an oscillating function with decaying amplitude.

The Lefschetz thimbles are a set of special submanifolds within the complexifed space that contain critical points of the action, and points on the thimble will flow to (or from) these critical points. Equivalently, Lefschetz thimbles are the manifolds that are generated by the gradient flow from critical points. When there are many critical points it can be difficult, in practise, to determine which thimbles should contribute to the integral. The Generalized Thimble Method takes a finite flow time from the original integration manifold to ${\cal M}$, and this manifold will approach the appropriate set of thimbles for the integral as the flow time is increased. Although this automatically selects the correct thimbles, in practise the numerical sampling algorithm can get stuck on one particular thimble, as the connections between the thimbles are exponentially small. This manifests itself as a multimodal problem in the Monte Carlo calculation of the integral when there is more than one thimble. Ideally then, one would prefer to work with systems that have a single thimble, and so a single critical point of the action.

We are interested in applying the thimble approach to real-time quantum systems, and in this context the critical points correspond to classical trajectories that extremize the action. The idea we shall follow, that allows us to work with a single thimble at a time, is that there is a single classical solution for a given initial position and velocity or, in the language of fields, a given $\varphi(t=0,\underline x)$ and $\dot\varphi(t=0,\underline x)$. Of course, given that we are studying a quantum system, there will be an ensemble of initial positions and velocities described by an initial density matrix, but we will see that we are able to separate the path integral into a two-step sampling procedure; for each member of the initial condition ensemble, we may compute a well-defined contribution to the path integral using the Generalized Thimble Method, and subsequently average over the initial condition ensemble in a straightforward way. 

The framework where one can separate the full path integral into an initial distribution and the subsequent dynamical part of the path integral already exists, and is known as the Schwinger-Keldysh, or in-in formalism \cite{Schwinger:1960qe,Keldysh:1964ud}. 
It is adapted to situations where one has initial data, rather than comparing in and out states, and as such one uses a time contour that starts at $t_0$, extends to some $T$, and then goes back to $t_0$, rather than to infinity. The value of $T$ is arbitrary, so long as the path encompasses any operators ${\cal O}(t)$ that one is interested in. For some theoretical situations it is useful to take $T\to\infty$, but for numerical simulations, such as in this paper, we work with finite $T$. We shall show in \cref{sec:free} that it is the same reasoning behind the freedom of choosing $T$ that enables us to separate the full path integral into two steps. 

We will see that our approach to solving for the complete real-time quantum dynamics may be linked to popular approximation schemes, such as the classical-statistical approximation, (truncations of) real-time Schwinger-Dyson (Kadanoff-Baym) equations \cite{Berges:2000ur} or a quantum ``dressing" of the classical path by Langevin methods in stochastic quantization \cite{Berges:2006xc}. As for traditional Euclidean equilibrium lattice simulations we may compute the path integrals exactly from first principles, up to lattice discretization errors and finite numerical resources. 

The structure of the paper is as follows:
in \cref{sec:thimble} we describe the Lefschetz Thimble Method and the Generalized Thimble Method, and introduce critical points and their role in evaluating the path integral. We connect to earlier work \cite{Alexandru:2017lqr}, and recall the flow equations to be used later on. 
In \cref{sec:real-time} we discretize the path integral and show how the initial conditions may be separated from the remaining degrees of freedom, and set up the two-step sampling procedure. We set up a convenient parametrization of the discretized path integral variables entering in the real-time, but not necessarily closed-time path, path integral. We demonstrate how splitting up the sum over paths into subsets with fixed initial conditions, can resolve the multimodal problem in a straightforward way. We then explicitly derive the Gaussian initial density matrix, at finite temperature and in the vacuum, and take care of some technical points that arise. 
In section \ref{sec:numerics} we present our numerical model and algorithm for a field theory in any dimension, and demonstrate our approach for a theory in 0+1 dimensions, so quantum mechanics.
We conclude in section \ref{sec:summary}. Some details of the perturbative one-loop correlator are placed in appendix \ref{app:loop}.

\section{The path integral deformed into the complex plane}
\label{sec:thimble}

Consider the path integral written in the form\footnote{We use the notation $\mathcal I$ as this connects with the standard literature (maybe up to a minus sign), but we ultimately have in mind that $\mathcal I = -\frac{iS}{\hbar}$, where $S$ is the action.},
\begin{align}
\label{eq:scalar}
\int_{{\mathbb R}^n} \prod_{i=1}^n \ud\varphi_i e^{-{\mathcal I}} ,
\end{align}
with real variables $\varphi_i$, and ${\mathcal I}$ is a function of all $\varphi_i$.
Here we combine space-time indices into $i$, and will specify them more precisely later.
As in the Feynman path integral, the exponent could be purely imaginary, so that the integrand is oscillatory with a constant amplitude.
We can improve the convergence of the integral through complexifying $\varphi_i$ and, because of Cauchy's theorem, we can deform the real integration cycle into the complex plane and still obtain the same result for the integral. In the following we shall use $\varphi_i$ to denote the real field, and $\phi_i$ shall denote the complexified field. As such, the initial integration manifold is ${\mathbb R}^n$, parametrized by $\varphi_i$. This integration cycle is then deformed to a surface in ${\mathbb C}^n$ with $n$ real dimensions, parametrized by $\phi_i$. 

\subsection{Lefschetz Thimble Method}

Such an approach is pioneered in \cite{Witten:2010cx,Witten:2010zr} for Feynman's path integral, with the altered integration cycle known as Lefschetz thimbles, obtained by gradient flow,
\begin{align}
\label{eq:GF}
\frac{\ud\phi_i}{\ud\tau} = \overline{\frac{\partial {\mathcal I}}{\partial \phi_i}},
\end{align}
 from critical points that are determined by $\left.\partial {\mathcal I}/\partial \phi_i\right|_{crit}=0$.
The over-line above refers to complex conjugation, and ${\mathcal I}$ is now considered a holomorphic function of the complex $\phi_i$.
The Lefschetz thimbles are $n$-dimensional integration cycles in the $n$-dimensional complex (so $2n$ real dimensional) plane.
As we can see from the flow equation, $\ud{\mathcal I}/\ud\tau=\sum_i|\partial {\mathcal I}/\partial \phi_i|^2$, ${\rm Im}[{\mathcal I}]$ is constant on each thimble, and of the same value as at the critical point, while ${\rm Re}[{\mathcal I}]$ keeps increasing with $\tau$ as we move away from the critical point, so its contribution to the integral (\ref{eq:scalar}) is exponentially suppressed away from the critical point.
As a result, we achieve quick convergence by performing the integral on the Lefschetz thimbles.

The idea of integrating over Lefschetz thimbles can be naturally adopted to numerical simulations \cite{Cristoforetti:2012su}, especially through Monte Carlo methods with, for example, Langevin dynamics \cite{Cristoforetti:2013wha,Aarts:2013fpa} and also Metropolis algorithms \cite{Mukherjee:2013aga,Alexandru:2015xva,Alexandru:2015sua}.
In the following sections, we will use the term Lefschetz Thimble Method to refer to the methods of generating samples on Lefschetz thimbles.
In the case of a single integration variable, the constraint that ${\rm Im}[{\mathcal I}]$ is the same as it is at the critical point can almost determine the thimbles entirely \cite{Aarts:2013fpa}.
With more integration variables, however, this one constraint is not sufficient and we should return to using the gradient flow (\ref{eq:GF}).
To be precise, we should consider the flow starting from a small neighbourhood of the critical point on each contributing thimble, as the gradient flow will actually take infinite time to run away from the critical point itself.
The neighbourhood should also be small enough to use an expansion of ${\mathcal I}$ up to quadratic terms, and
with only these quadratic terms present we can solve the flow equation explicitly.
This requires each isolated critical point, $p$, to be non-degenerate \cite{Cristoforetti:2012su},
\begin{align}
\left.\frac{\partial {\mathcal I}}{\partial \phi_i}\right|_{p} =0,
\quad {\rm  and} \quad
{\rm det}\left.\left(\frac{\partial^2 {\mathcal I}}{\partial \phi_i\partial \phi_j} \right)\right|_p\neq0.
\end{align}
By  Morse theory/Picard-Lefschetz theory, the matrix of second order derivatives of ${\rm Re}[{\mathcal I}]$ has $n$ positive eigenvalues and $n$ negative ones, and
near the critical point, we can approximate the Lefschetz thimble with the manifold generated by these $n$ positive eigenvalues/eigenvectors.

The ``sign problem'' is milder on the Lefschetz thimbles than on the real space.
On each thimble, ${\rm Im}[{\mathcal I}]$ is constant, and the only varying complex phase comes from the Jacobian of the transformation that maps the complex integration variables into real ones \cite{Mukherjee:2013aga}.
More importantly, the exponential suppression of the magnitude away from the critical point makes the Monte Carlo simulation on each thimble possible.

A subtlety arises when there exists multiple critical points since
in this case one has to find all the critical points and related Hessian matrices analytically, and then decide which combination of thimbles is equivalent to the original integration contour.
There could exist one or more dominant critical points, giving similar contributions to the path integral.
But there also exist concrete examples where equally dominant critical points cancel each other out in the integral, so that the main contribution comes from sub-dominant critical points \cite{Behtash:2015kna,Dunne:2015eaa}.
One might also have to sum over all contributing thimbles to not miss something \cite{Tanizaki:2015rda}.
This is not an easy task for a general theory,
so is there a technique that includes the complete integration cycle automatically, without having a ``sign problem'' at the same time?
Such a technique is the Generalized Thimble Method.

\subsection{Generalized Thimble Method}
\label{sec:GenThimbles}

The gradient flow (\ref{eq:GF}) serves two purposes.
On the one hand, starting near critical points, it defines the corresponding Lefschetz thimbles.
On the other hand, it maps the real integration cycle to the combination of thimbles contributing to the original integral.
For instance, at $\tau=0$, we have the original $n$-dimensional real space, and as $\tau\to+\infty$ we obtain the right combination of Lefschetz thimbles.
In fact, the flow equation (\ref{eq:GF}) in this case generates a family of $n$-manifolds that are characterized by the flow time, $\tau$, and at any such flow time the integral would return the same result.
Given the ``sign problem'' at $\tau=0$ and its absence at $\tau=+\infty$, one might expect the ``sign problem'' to be alleviated gradually along $\tau$, and even at some finite $\tau$ the Monte Carlo simulation may already become effective.
This turns out to be the case and such a finite $\tau$ approach, which is known as Generalized Thimble Method \cite{Alexandru:2017czx,Alexandru:2018fqp,Alexandru:2018ngw}, has many applications in dealing with the ``sign problem'' in different scenarios \cite{Alexandru:2016gsd,Alexandru:2017lqr,Alexandru:2017oyw,Alexandru:2017czx,Alexandru:2018fqp,Alexandru:2018ngw}.

The finite $\tau$ manifold, ${\mathcal M}$, has $n$ real dimensions and is embedded in an $n$-dimensional complex plane.
We can parametrize it with real variables as follows.
Provided with initial real values $\varphi_i$, the flow equation (\ref{eq:GF}) transforms the fields into complex $\phi_i$.
Thus we arrive at the equalities,
\begin{align}
\int_{{\mathbb R }^n}\prod_{i=1}^n\ud\varphi_i e^{-{\mathcal I}(\varphi)}=
\int_{{\mathcal M}}\prod_{i=1}^n\ud\phi_i e^{-{\mathcal I}(\phi)}=
\int_{{\mathbb R}^n}\prod_{i=1}^n\ud\varphi_i {\rm det}\left(\frac{\partial \phi}{\partial \varphi}\right)e^{-{\mathcal I}(\phi(\varphi))}.
\end{align}
The first equality is where we complexify $\varphi_i\to\phi_i$ and perform the integration over the manifold ${\cal M}$\footnote{Note that for zero flow-time, ${\cal M}$ is just the initial real manifold ${\mathbb R}^n\subset{\mathbb C}^n$.}; the second equality is where we think of $\phi_i(\tau_{final})$ as a function of the initial $\varphi_i=\phi_i(\tau=0)$, and perform a co-ordinate transformation back to $\varphi_i$. Note that in the final expression, ${\mathcal I}$ is evaluated at $\phi_i(\varphi)$, while the first expression is evaluated at $\varphi_i$, with the Jacobian providing the appropriate correction factor.

We can also deduce from the flow equation (\ref{eq:GF}) that the Jacobian matrix $J_{ij}=\partial \phi_i /\partial \varphi_j$ satisfies 
\begin{align}
\frac{\ud}{\ud\tau}\left(\frac{\partial \phi_i}{\partial \varphi_j}\right) = \overline{\frac{\partial^2 {\mathcal I}}{\partial \phi_i\partial \phi_k}\frac{\partial \phi_k}{\partial \varphi_j}},
\end{align}
with $J_{ij}$ an $n\times n$ identity matrix at $\tau=0$.
In practice, one can carry out importance sampling with the weight $P(\varphi)=e^{-{\rm Re}[{\mathcal I}]+\ln|{\rm det}(J)|}$, and then reweight by the remaining imaginary parts,
\begin{align}
\label{eq:reweighting}
\langle {\mathcal O}(\phi)\rangle = 
\frac{\Big\langle e^{-i{\rm Im}[{\mathcal I}]+i{\rm arg}\left({\rm det}(J)\right)} {\mathcal O}(\phi)\Big\rangle_P }{\Big\langle e^{-i{\rm Im}[{\mathcal I}]+i{\rm arg}\left({\rm det}(J)\right)}\Big\rangle_P }
.
\end{align}
We see this by noting that  expectation values for operators are given by the following path integral
\begin{align}
\langle{\cal O}\rangle&\sim\int_{{\mathbb R}^n} \prod_{i=1}^n \ud\varphi_i \;{\cal O}(\varphi)e^{-{\mathcal I}}
	\sim\int_{{\mathbb R}^n}\prod_{i=1}^n\ud\varphi_i {\rm det}\left(J\right)\;{\cal O}(\phi(\varphi))e^{-{\mathcal I}(\phi(\varphi))}\\\nonumber
	&\sim\int_{{\mathbb R}^n}\prod_{i=1}^n\ud\varphi_i \;{\cal O}(\phi(\varphi))e^{-i{\rm Im}[{\mathcal I}]+i{\rm arg}\left({\rm det}(J)\right)}e^{-{\rm Re}[{\mathcal I}]+\ln|{\rm det}(J)|}.
\end{align}

While the Lefschetz Thimble Method approach is well-suited to an analytic approach, the Generalized Thimble Method with finite $\tau$ is more numerically oriented.
On the other hand, the Generalized Thimble Method is not sensitive to the degeneracy of critical points.

To alleviate the ``sign problem'' one may have to go to a manifold with large $\tau$, where the connection among different regions of the integration contour, flowing from multiple critical points, becomes exponentially small. As a result, simple Monte-Carlo sampling algorithms may get stuck in one region. This ``multimodal" problem is a likely feature of the Generalized Thimble Method.
Many sophisticated methods have been proposed to get the correct exploration of the manifold \cite{Alexandru:2017oyw,Fukuma:2017fjq}.
But there is no doubt that both the Lefschetz Thimble Method and the Generalized Thimble Method are effective in the case of a single critical point.
Then a natural question is whether we can tell the number of critical points beforehand.
It turns out that we can, at least for a scalar theory.

\section{Theoretical developments for the real-time path integral}
\label{sec:real-time}

At this point, we will derive a series of results for the path integral, which will all come into play, when we put together our algorithm in section \ref{sec:numerics}. 

\subsection{The path integral}
\label{sec:PathInt}

To fix our conventions we will start by deriving the path integral expression for calculating operator expectation values in the Heisenberg picture, $\langle\hat{\mathcal{O}}\big(\hat\Phi,\hat\Pi\big)\rangle$, with operator $\hat{\mathcal{O}}$ consisting of the scalar field operator $\hat\Phi$ and its canonical conjugate, $\hat\Pi$, at one or more times.
We follow the convention of \cite{Weinberg:1995mt}
:
\begin{align}
\int {\mathcal D}\phi |\phi;t \rangle \langle \phi;t| = 1
,~~
\int {\mathcal D}\pi |\pi;t \rangle \langle \pi;t| = 1
,~~
\langle \phi;t |\pi;t\rangle = \left[\frac{\ud^dx}{2\pi\hbar }\right]^{\frac{(N_s)^d}{2}} \exp\left(\frac{i}{\hbar}\int \ud^dx \pi(x)\phi(x)\right),
\end{align}
where $|\phi;t\rangle$ and $|\pi;t\rangle$ are eigenvectors of operator $\hat\Phi(t)$ and $\hat\Pi(t)$ respectively.
In the formulae above, a discretized $d$-dimensional space was assumed. That is, $N_s$ sites along each spatial direction and distance $\ud x$ between two neighbouring sites, so the volume $V=(N_s\ud x)^d$  and, furthermore, we suppressed the spatial index. For instance, ${\mathcal D}\phi=\prod_x \ud\phi(x)$.
It is also convenient to switch between continuous and discrete expressions via,
\begin{align}
\int \ud^dx \quad \Leftrightarrow \quad \sum_x \ud^dx
,\qquad\qquad
\frac{\delta}{\delta \phi(x)} \quad \Leftrightarrow \quad \frac{1}{\ud^dx}\frac{\partial}{\partial \phi(x)} .
\end{align}

\begin{figure}[h]
\centering
\includegraphics[width=0.65\textwidth]{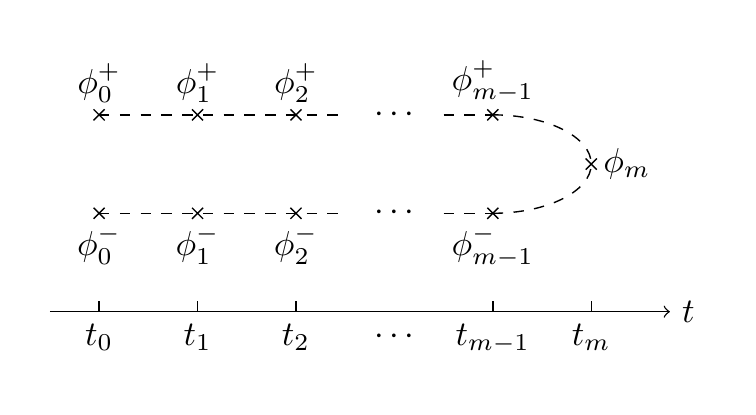}
\caption{Illustration of inserted $\phi$. Here we separate $\phi^+$ and $\phi^-$ vertically for demonstration purpose.  All these fields live on the real-time line. The difference between two neighbouring $t$ is a constant, given by $\ud t$.}
\label{fig:seq}
\end{figure}

We can then calculate $\langle\hat{\mathcal{O}}\big(\hat\Phi,\hat\Pi\big)\rangle$  by inserting complete sets of $ | \phi;t_i\rangle\langle \phi;t_i|$ in succession along the temporal direction, leading to
\begin{align}
\label{eq:operator}
&\qquad\qquad \qquad\qquad \langle \hat{\mathcal{O}}\big(\hat\Phi,\hat\Pi\big) \rangle = {\rm Tr} \left[ \hat{\mathcal{O}}\big(\hat\Phi,\hat\Pi\big) \hat{\rho}\big(\hat\Phi(t_0),\hat\Pi(t_0)\big) \right]
=
\\
&\int{\mathcal D}\phi
\langle \phi_0^-;t_0 
|\phi_{1}^-;t_{1}\rangle \langle \phi_{1}^-;t_{1} | 
\cdots
\hat{\mathcal{O}}
\cdots
|\phi_1^+;t_1\rangle \langle \phi_1^+;t_1 
|\phi_0^+;t_0\rangle \langle \phi_0^+;t_0 | 
~\hat{\rho}~
|\phi_0^-;t_0\rangle 
,
\nonumber
\end{align}
with $\hat{\rho}\big(\hat\Phi(t_0),\hat\Pi(t_0)\big)$ the initial density matrix operator at $t_0$.
\Cref{fig:seq} gives a graphic demonstration of the insertion along the temporal direction.

In the presence of operators $\mathcal{O}$, the insertion is not unique.
There are two features worth noting.
(1) There are different ways for the operators to appear in the expression.
For instance, in the case of $\hat{\mathcal O}=\hat\Phi(t_\alpha)\hat\Phi(t_\beta)$ and $t_\alpha>t_\beta$, if $\hat\Phi(t_\beta)$ appears in the upper ($\phi^+$) layer, then $\hat\Phi(t_\alpha)$ can appear either in the upper ($\phi^+$) or lower ($\phi^-$) layer.
We will see what this implies for the path integral in section \ref{subsec:feynman}.
(2) One is free to choose the turning point $\phi_m$, as long as the contour includes the operator $\hat{\mathcal O}$.
The path integrals with different turning points give the same expectation value of the operator.

First, we need to calculate each Feynman kernel $\langle \phi_{i};t_{i} |\phi_{j};t_{j}\rangle$.
Here we only assume that the time difference $|t_i-t_j|$ is small, but do not specify which time is earlier.
Since we want to derive the path integral with $\ud t$ finite, a symmetric expression of the kernel seems a better choice, as it will converge more quickly in the limit $\ud t\to 0$.
Thus by evolving each state to the equal time $\tilde t=(t_i+t_{j})/2$, and then inserting the complete set of $|\pi;\tilde t\rangle \langle \pi;\tilde t |$, we arrive at the expression,
\begin{align}
\label{eq:kernel}
&\langle \phi_{i};t_{i} |\phi_{j};t_{j}\rangle 
=\langle \phi_{i};\tilde t| \exp\left(-\frac{i}{\hbar} \frac{t_{i}-t_{j}}{2} \hat{H}\right)  \exp\left(-\frac{i}{\hbar} \frac{t_{i}-t_{j}}{2} \hat{H}\right)  |\phi_{j};\tilde t\rangle 
\nonumber\\
=&\int {\mathcal D}\pi
\langle \phi_{i};\tilde t| \exp\left(-\frac{i}{\hbar} \frac{t_{i}-t_{j}}{2} \hat{H}\right) |\pi;\tilde t\rangle \langle \pi;\tilde t| \exp\left(-\frac{i}{\hbar} \frac{t_{i}-t_{j}}{2} \hat{H}\right)  |\phi_{j};\tilde t\rangle 
\nonumber\\
=&\left[\frac{\ud^dx}{2\pi\hbar }\right]^{(N_s)^d} \int {\mathcal D}\pi
\exp\left(-\frac{i}{\hbar}(t_{i}-t_{j})\frac{H\big(\phi_{i},\pi\big)+H\big(\phi_{j},\pi\big)}{2}+\frac{i}{\hbar}\int \ud^dx \pi(\phi_{i}-\phi_{j})\right) 
\nonumber\\
=&\left[\frac{\ud^dx}{i2\pi\hbar(t_i-t_{j}) }\right]^{\frac{(N_s)^d}{2}}  \exp\left(\frac{i}{\hbar}(t_{i}-t_{j})L\big(\phi_{i},\phi_{j}\big)\right) 
,
\end{align}
where the operator $\hat{H}$ is the Hamiltonian, which contains only up to quadratic terms of $\hat\Pi$.  
For the scalar theory, we assume the general expression,
\begin{align}
\hat{H}=\int \ud^dx\left(\frac{\hat\Pi^2}{2} +\hat{C}(\hat\Phi)\right)
,
\end{align}
with $\hat{C}(\hat\Phi)$ composed of spatial derivative terms and a field potential.
We do not need to know the exact expression of  $\hat{C}(\hat\Phi)$ at the moment, but demand $\hat{C}(\hat\Phi)$ is local in time. We also assume that all operators, for instance $\hat{H}$, may be written as functions of variables $\phi$ and $\pi$.
The function $H(\phi_i,\pi)$ is then the result of the operator $\hat{H}$ acting on states. 

Given the Hamiltonian, the Lagrangian is
\begin{align}
L\big(\phi_i,\phi_{j}\big)=\int \ud^dx\left(\frac{1}{2}\left[\frac{\phi_i(x)-\phi_{j}(x)}{t_i-t_{j}}\right]^2 -\frac{C(\phi_i)+C(\phi_{j})}{2}\right).
\end{align}
which is symmetric in $i$ and $j$.
In light of \cref{eq:kernel}, the wave function $\langle \phi;t  |{\rm in}\rangle =\int {\mathcal D}\phi' \langle \phi;t |\phi';t-\ud t\rangle \langle \phi';t-\ud t |{\rm in}\rangle $ satisfies the Schr\"odinger functional equation \cite{Jackiw:1988sf},
\begin{align}
\label{eq:schrodinger}
i\hbar \frac{\partial}{\partial t} \langle \phi,t | {\rm in} \rangle = \int \ud^dx \left[-\frac{\hbar^2}{2}\frac{\delta^2}{\delta \phi(t,x)^2} + C(\phi(t,x))\right] \langle \phi,t | {\rm in} \rangle,
\end{align}
in the limit $\ud t\to 0$.
Thus, Feynman's kernel is the propagator for small time intervals.
We emphasize that the derivations in (\ref{eq:kernel}) are valid for both $t_i>t_{j}$ and $t_i<t_{j}$.
From $\phi_{m-1}^+$ to $\phi_m$, the time difference is $\ud t$, but from  $\phi_m$ to $\phi_{m-1}^-$, it is $-\ud t$.

Now we can continue working with \cref{eq:operator} 
\begin{align}
&\qquad\qquad  \langle \hat{\mathcal{O}}\big(\hat\Phi,\hat\Pi\big) \rangle = {\rm Tr} \left[ \hat{\mathcal{O}}\big(\hat\Phi,\hat\Pi\big) \hat{\rho}\big(\hat\Phi(t_0),\hat\Pi(t_0)\big) \right]
\nonumber\\
=&~{\mathcal N}\int {\mathcal D}\phi
\exp\left(\frac{i}{\hbar}\int_{\mathcal C} \ud t L\right)
\mathcal{O}\big(\phi,\pi\big)
 \langle \phi^+_0;t_0 | 
\hat{\rho}\Big(\hat\Phi(t_0),\hat\Pi(t_0)\Big)
|\phi_0^-;t_0\rangle 
,
\end{align}
where ${\mathcal N}$ is a collection of numerical constants that appear in kernel (\ref{eq:kernel}), and the integration contour ${\mathcal C}$ is understood as the contour shown in \cref{fig:seq}.
In the discrete theory, the integral over ${\mathcal C}$ in the exponent is really an abbreviation of,
\begin{align}
\int_{\mathcal C} \ud t L = \ud t\sum_{i=1}^{m} \left(L\big(\phi_i^+,\phi_{i-1}^+\big) - L\big(\phi_{i-1}^-,\phi_{i}^-\big)\right)
,
\end{align}
where, to write the expression elegantly, we denote $\phi_m=\phi_m^+=\phi_m^-$.
On the other hand, since the numerical constant ${\mathcal N}$ does not depend on the operator $\hat{{\mathcal O}}$, we can fix it by taking the case $\hat{{\mathcal O}}=1$,
\begin{align}
\label{eq:o1}
1={\rm Tr}\left[\hat{\rho}\big(\hat\Phi(t_0),\hat\Pi(t_0)\big) \right]={\mathcal N}\int {\mathcal D}\phi
\exp\left(\frac{i}{\hbar}\int_{\mathcal C} \ud t L\right)
 \langle \phi_0^+;t_0 | 
\hat\rho\Big(\hat\Phi(t_0),\hat\Pi(t_0)\Big)
|\phi_0^-;t_0\rangle ,
\end{align}
where we utilize the fact that the trace of the density matrix is one.
Therefore, we can write the expectation value of the operator as, 
\begin{align}
\label{eq:expectation1}
 \langle \hat{\mathcal{O}}\big(\hat\Phi,\hat\Pi\big) \rangle = \frac{\int {\mathcal D}\phi
\exp\left(\frac{i}{\hbar}\int_{\mathcal C} \ud t L\right)
\mathcal{O}\big(\phi,\pi\big)
 \langle \phi_0^+;t_0 | 
\hat\rho\Big(\hat\Phi(t_0),\hat\Pi(t_0)\Big)
|\phi_0^-;t_0\rangle }
{\int {\mathcal D}\phi
\exp\left(\frac{i}{\hbar}\int_{\mathcal C} \ud t L\right)
 \langle \phi_0^+;t_0 | 
\hat\rho\Big(\hat\Phi(t_0),\hat\Pi(t_0)\Big) 
|\phi_0^-;t_0\rangle }.
\end{align}
We will compute
\Cref{eq:expectation1} by 
a Monte Carlo evaluation, where one generates samples according to the distribution in the denominator,
\begin{align}
\label{eq:denominator}
\int {\mathcal D}\phi
\exp\left(\frac{i}{\hbar}\int_{\mathcal C} \ud t  L\right)
 \langle \phi_0^+;t_0 | 
\hat{\rho}\Big[\Phi(t_0),\Pi(t_0)\Big] 
|\phi_0^-;t_0\rangle.
\end{align}

\subsection{Critical points}
\label{sec:critical_points}

We are now in a position to find the critical points in \cref{eq:denominator}.
We write \mbox{${\mathcal I}=-i\int_{\mathcal C} \ud t L/\hbar + \cdots$}, with ellipsis denoting extra terms coming from the initial density matrix, which are only functions of $\phi_0^+$ and $\phi_0^-$.
To study the critical points it is convenient to use another basis, $\phi^{cl}$ and $\phi^{q}$, defined through \cite{Schwinger:1960qe,Keldysh:1964ud,Greiner:1996dx,Aarts:1997kp,Kamenev:2009}\footnote{
In the literature, there exist alternative ways to transform $\phi^+$ and $\phi^-$, with 
Keldysh's original convention \cite{Keldysh:1964ud,Kamenev:2009} corresponding to $\phi^{\pm}=\left( \phi^{cl}\pm \phi^q\right)/\sqrt{2}$.
Here we follow the approach of \cite{Greiner:1996dx,Aarts:1997kp}, but we adopt the names $\phi^{cl}$ and $\phi^q$ from \cite{Kamenev:2009}.
}, 
\begin{align}
\phi_i^+(x)=\phi_i^{cl}(x)+\frac{\phi_i^q}{2}(x)
,\quad
\phi_i^-(x)=\phi_i^{cl}(x)-\frac{\phi_i^q}{2}(x).
\end{align}

With these\footnote{Even though we do not apply the change of basis to $\phi_m(x)$, as there is only one field, it will be useful to introduce $\phi_m^{cl}(x)=\phi_m(x)$ and $\phi_m^{q}(x)=0$. But we do not treat $\phi_m^q(x)$ as a variable.}, the action becomes
\begin{align}
\label{eq:L}
\int_{\mathcal C} \ud t L = \ud t\sum_{i=1}^{m} \left[\int \ud^dx\left(\frac{\left(\phi_i^{cl}(x)-\phi_{i-1}^{cl}(x)\right)\left(\phi_i^{q}(x)-\phi_{i-1}^{q}(x)\right)}{\ud t^2}
\right)-\frac{E_i+E_{i-1}}{2}\right]
,
\end{align}
where
\begin{align}
\label{eq:Ei}
E_i=\int \ud^dx \left[C\left(\phi_i^{cl}(x)+\frac{\phi_i^q}{2}(x)\right)-C\left(\phi_i^{cl}(x)-\frac{\phi_i^q}{2}(x)\right)\right]
.
\end{align}
We may derive two general results without knowing the explicit form of the Lagrangian:

1. $E_m=0$. 
The only term  in the exponent containing $\phi_m(x)$ is the product of $\phi_m(x)$ and $\phi_{m-1}^q(x)$.
Actually, in \cref{eq:denominator}, one can integrate $\phi_m(x)$ out, and get a delta function, as follows,
\begin{align}
\label{eq:integral1}
 \int {\mathcal D}\phi_m e^{-\frac{i}{\hbar \ud t}\int \ud^dx\phi_m(x)\phi_{m-1}^q(x)}
=
 \prod_x(2\pi)\delta\left(-\frac{\ud^dx}{\hbar \ud t}\phi_{m-1}^q(x)\right)
=
\left(\frac{2\pi\hbar \ud t}{\ud^dx}\right)^{(N_s)^d} \prod_x\delta\left(\phi_{m-1}^q(x)\right).
\end{align}
If one further integrates out $\phi_{m-1}^q(x)$, \cref{eq:denominator} would become the same form as the original integral, but with the turning point $\phi_m$ replaced by $\phi_{m-1}^{cl}$, and with an extra overall constant.
We emphasize the fact that the integration over $(\phi_{m}^{cl},~\phi_{m-1}^q)$ together is a constant, and it will not alter the remaining path integral, except through the overall constant.
One may integrate out the $(\phi_{i}^{cl},~\phi_{i-1}^q)$ one by one, as they become the last pair along the real-time direction.
By continuing this process down to $\phi_0$, we arrive at 
\begin{align}
\int {\mathcal D}\phi
\exp\left(\frac{i}{\hbar}\int_{\mathcal C} \ud t L\right)
 \langle \phi_0;t_0 | 
\hat{\rho}
|\phi_n;t_0\rangle
=\cdots=\frac{1}{\mathcal N}\int {\mathcal D}\phi
 \langle \phi_0;t_0 | 
\hat{\rho} 
|\phi_0;t_0\rangle
=\frac{1}{\mathcal N}
.
\end{align}
This is just \cref{eq:o1}, written in reverse order, and also provides an alternative way to compute the constant ${\mathcal N}$.
Of course, to avoid keeping numerical constants, one can execute such contraction simultaneously in both the numerator and denominator of \cref{eq:expectation1}.
However, the contraction in the numerator is no longer valid once $\hat{\mathcal O}(t)$ is reached.
Generally, if $t_{max}$ is the maximum time that the operator $\hat{\mathcal O}$ depends on, then as long as $t_m\geqslant t_{max}$, the path beyond $t_{max}$ is contractible.
This corresponds to the freedom that one can have in choosing the closed time path when restricted to the real-time line.

2. All terms in $E_i$ contain odd powers of $\phi_i^{q}(x)$, as even powers of $\phi_i^{q}(x)$ cancel out.
One can check this by expanding \cref{eq:L,eq:Ei} as a Taylor series in $\phi_i^q$.
In fact, the quantum field theory can be computed in perturbation theory of $\phi^q$ \cite{Aarts:1997kp}.
The leading order theory has a term linear in $\phi^q$ appearing in the exponent, and if we carry out the integration of $\phi^q$ explicitly, the leading order theory is the classical theory.
A simple example is $\lambda\phi^4$ theory (suppressing for moment the initial density matrix part of the expression), 
\begin{align}
&\int {\mathcal D}\phi \exp\left(\frac{i}{\hbar}\int_{\mathcal C}\ud t\int \ud^dx\left[\frac{1}{2}\left(\dot\phi\right)^2-\frac{1}{2}\left(\nabla\phi\right)^2-\frac{1}{2}m^2\phi^2-\frac{\lambda}{4!}\phi^4\right]\right)
\nonumber \\
= &
\int {\mathcal D}\phi \exp\left(\frac{i}{\hbar}\int \ud t\int \ud^dx\left[\dot\phi^{cl}\dot\phi^{q}-\nabla\phi^{cl}\nabla\phi^{q}-m^2\phi^{cl}\phi^{q}-\frac{\lambda}{4!}\left[4\phi^{q}(\phi^{cl})^3+(\phi^{q})^3\phi^{cl}\right]\right]\right)
\nonumber \\
= &
\int {\mathcal D}\phi e^{\frac{i}{\hbar}\int \ud t\int \ud^dx\left[\dot\phi^{cl}\dot\phi^{q}-\nabla\phi^{cl}\nabla\phi^{q}-m^2\phi^{cl}\phi^{q}-\frac{\lambda}{3!}\phi^{q}(\phi^{cl})^3\right]}\left(1-\frac{i\lambda}{4!\hbar}\int \ud t\int \ud^dx(\phi^{q})^3\phi^{cl}+\cdots\right)
,
\nonumber
\end{align}
By keeping the leading term in the final factor, and then integrating out $\phi^q$, we find the delta function,
\begin{align}
\delta\left(-\frac{\partial^2 \phi^{cl}}{\partial t^2} 
+\nabla^2 \phi^{cl}
-m^2\phi^{cl}
-\frac{\lambda}{3!}\left(\phi^{cl}\right)^3\right),
\end{align}
which means, in the leading order theory, that $\phi^{cl}$ satisfies the equation of motion of the classical field.
More generally, $\frac{\partial {\mathcal I}}{\partial \phi^q}\Big|_{\phi^q=0}=0$ leads to the classical equation of motion.
Furthermore, when $\phi_i^q(x)=0$ at any $x$, then $\partial E_i/\partial \phi_i^{cl}(x)$ must also vanish, since it consists of odd terms of $\phi_i^q$. \\

We may write down straightforwardly for $0< i< m$,
\begin{align}
\label{eq:deriv1}
\frac{\partial {\mathcal I}}{\partial \phi_i^q(x)}
&=-\frac{i(\ud t)(\ud^dx)}{\hbar}\left[
\frac{2\phi_{i}^{cl}(x)-\phi_{i-1}^{cl}(x)-\phi_{i+1}^{cl}(x)}{(\ud t)^2}
-\frac{\partial E_i}{\partial  \phi_i^q(x)}
\right]
,
\\
\label{eq:deriv2}
\frac{\partial {\mathcal I}}{\partial \phi_i^{cl}(x)}
&=-\frac{i(\ud t)(\ud^dx)}{\hbar}\left[\frac{2\phi_{i}^q(x) -\phi_{i-1}^{q}(x)-\phi_{i+1}^{q}(x)}{(\ud t)^2}
-\frac{\partial E_i}{\partial  \phi_i^{cl}(x)}
\right]
,
\end{align}
and for $i=m$,
\begin{align}
\label{eq:deriv3}
\frac{\partial {\mathcal I}}{\partial \phi_m(x)}
&=\frac{i(\ud^dx)}{\hbar \ud t}\phi_{m-1}^{q}(x)
.
\end{align}
We now note that the critical points are determined by $\left.\partial {\mathcal I}/\partial \phi\right|_{crit}=0$ for all $\phi$, from which it follows that \cref{eq:deriv1,eq:deriv2,eq:deriv3} all vanish at those points.
We can now show by induction, that critical points require all $\phi_i^{q}(x)=0$ with $0<i<m$.
This is true for $i=m-1$, as  the vanishing \cref{eq:deriv3} alone indicates $\phi_{m-1}^{q}(x)=0$ at any $x$.
Furthermore, if $\phi_{i+1}^{q}(x)=0$ along with $\phi_{i}^{q}(x)=0$ at any $x$, then as this implies $\partial E_i/\partial \phi_i^{cl}(x)=0$, we see that the vanishing of \cref{eq:deriv2} leads to $\phi_{i-1}^{q}(x)=0$.
We can apply this induction down to $\partial {\mathcal I}/\partial \phi_2^{cl}=0$, such that all  $\phi_i^{q}(x)=0$ with $0<i<m$. 

Now that we have $\phi_i^{q}(x)=0$ at the critical point, we can use the vanishing of \cref{eq:deriv1}, i.e. $\partial {\mathcal I}/\partial \phi_i^q(x)=0$, to lead us to the classical equation of motion,
\begin{align}
\label{eq:classic}
\frac{2\phi_{i}^{cl}(x)-\phi_{i-1}^{cl}(x)-\phi_{i+1}^{cl}(x)}{(\ud t)^2}
-\frac{\partial E_i}{\partial  \phi_i^q(x)} \Big|_{\phi_i^q=0}
=0.
\end{align}
Notice that the second term on the left-hand side contains only $\phi_i^{cl}$.
Therefore, \cref{eq:classic} determines $\phi_{i+1}^{cl}(x)$ uniquely once  $\phi_i^{cl}(x)$ and $\phi_{i-1}^{cl}(x)$ are known.
In other words, once $\phi_{0}^{cl}(x)$ and $\phi_{1}^{cl}(x)$ are known, we can uniquely solve all subsequent $\phi^{cl}$. 
In this sense, we can assert that the critical points are completely determined by $\phi_{0}^{cl}(x)$ and $\phi_{1}^{cl}(x)$.

What we have shown, therefore, is that there is a single critical point for each given $\phi_{0}^{cl}(x)$ and $\phi_{1}^{cl}(x)$, and so by picking $\phi_{0}^{cl}(x)$ and $\phi_{1}^{cl}(x)$ there will be a single thimble associated to that single critical point.
We now need a scheme to select $\phi_{0}^{cl}(x)$ and $\phi_{1}^{cl}(x)$, and for this we need an explicit expression of the initial density matrix.

\subsection{Thermal initial density matrix for a free field}
\label{sec:thermal}

Later on, we will be particularly interested in Gaussian initial conditions, which may then be chosen to be vacuum, thermal equilibrium or any out-of-equilibrium initial Gaussian state. 

But before we specialise to Gaussian states, we will first recall how a general thermal equilibrium state may be introduced as a path integral of imaginary time.

\begin{figure}[t]
\centering
\includegraphics[width=1.0\textwidth]{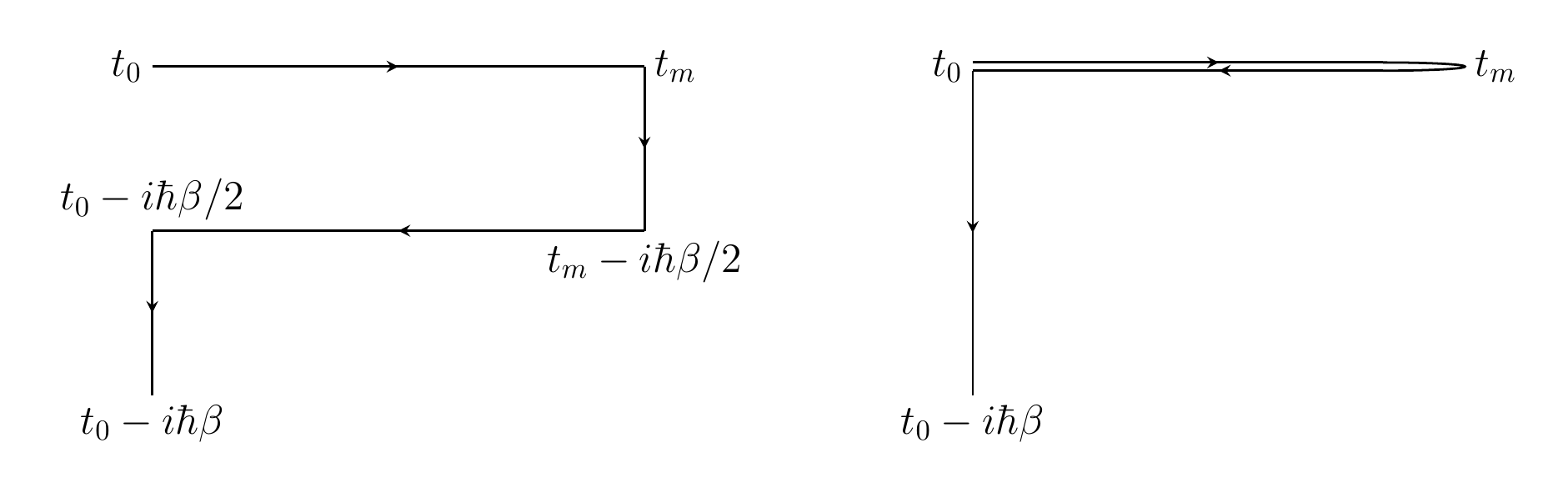}
\caption{
For thermal equilibrium, the complex time path is periodic along the imaginary time direction, with the period $\hbar\beta$, and
there is some freedom in choosing the contour in the complex time plane. 
(L) The  Schwinger-Keldysh closed time contour used in \cite{Alexandru:2016gsd,Alexandru:2017lqr};
(R) The Schwinger-Keldysh closed time contour used in \cite{Aarts:1997kp}.
In section \ref{sec:thermal} we use the right-hand side path to derive analytic expressions, where both trajectories of $t_0\to t_m$ and $t_m\to t_0$ are located on the real-time line, and the vertical offset between them exists only for demonstration purpose.
}
\label{fig:contour}
\end{figure}

The density matrix operator for thermal equilibrium is $\hat{\rho}=e^{-\beta \hat{H}}/Z$, where $1/\beta=k_BT$, with $k_B$ being Boltzmann's constant and $T$ the temperature. The normalization $Z={\rm Tr}\left[e^{-\beta \hat{H}}\right]$ is just an overall constant, which we will suppress for now. In this case, the insertion of complete sets leads to, 
\begin{align}
&\qquad \qquad \qquad \langle \phi_0^+;t_0 | 
e^{-\beta\hat{H}}
|\phi_0^-;t_0\rangle 
=
 \langle \phi_0^+;t_0 | 
e^{-\ud\beta \hat{H}}
\cdots
e^{-\ud\beta \hat{H}}
|\phi_0^-;t_0\rangle 
\\
=&
\int\prod_{k=I}^{N-1} {\mathcal D}\phi_k
\langle \phi_0^+;t_0 | e^{-\ud\beta \hat{H}}
|\phi_{I};t_0\rangle \langle \phi_{I};t_0 | e^{-\ud\beta \hat{H}}
|\phi_{II};t_0\rangle \langle \phi_{II};t_0 | 
\cdots
\langle \phi_{N-1};t_0 | e^{-\ud\beta \hat{H}}|\phi_0^-;t_0\rangle 
,
\nonumber
\end{align}
with $\ud\beta=\beta/N$.
As the label suggests, it would be convenient to also denote $\phi_0^+$ as $\phi_0$ and $\phi_0^-$ as $\phi_N$.
The computation of each single kernel is similar to (\ref{eq:kernel}), and we can also compute it in a symmetric way,
\begin{align}
&\langle \phi_{k};t_{0}| e^{-\ud\beta \hat{H}} |\phi_{k+1};t_{0}\rangle 
=
\int {\mathcal D}\pi \langle \phi_{k};t_{0}| e^{-\frac{\ud\beta}{2} \hat{H}}
|\pi_{k};t_{0}\rangle \langle \pi_{k};t_{0} | e^{-\frac{\ud\beta}{2} \hat{H}}
 |\phi_{k+1};t_{0}\rangle 
\nonumber\\
=&\left[\frac{d^dx}{2\pi\hbar }\right]^{(N_s)^d} 
\int {\mathcal D}\pi
\exp\left(-\ud\beta \frac{H\big[ \phi_k, \pi\big]+H\big[ \phi_{k+1}, \pi\big]}{2}+\frac{i}{\hbar}\int \ud^dx \pi(\phi_{k}-\phi_{k+1})\right)
\nonumber\\
=&\left[\frac{\ud^dx}{2\pi\hbar^2\ud\beta }\right]^{\frac{(N_s)^d}{2}} 
\exp\left(\ud\beta L\big[ \phi_{k},\phi_{k+1} \big]\right),
\label{eq:kernel_thermal}
\end{align}
where the Lagrangian is defined similarly to the real-time one, but with $\ud t$ substituted by $-i\hbar \ud\beta$,
\begin{align}
L\big[\phi_{k},\phi_{k+1}\big]
=\int \ud^dx\left[\frac{1}{2}\left(\frac{\phi_k(x)-\phi_{k+1}(x)}{-i\hbar \ud\beta}\right)^2 -\frac{C(\phi_k)+C(\phi_{k+1})}{2}\right].
\end{align}
It is then straightforward to compose the expectation value as a series of integrals, along a trajectory from $\phi_N$ (so $\phi_0^-$) to $\phi_0$ (so $\phi_0^+$), through negative imaginary time,
\begin{align}
\label{eq:density}
 &\langle \phi_0;t_0 | 
e^{-\beta \hat{H}}
|\phi_N;t_0\rangle 
=\left[\frac{\ud^dx}{2\pi\hbar^2\ud\beta }\right]^{\frac{N(N_s)^d}{2}}
\int\prod_{k=I}^{N-1} {\mathcal D}\phi_k
\exp\left(\ud\beta L_0\big[\phi_{0},\phi_{I}\big] \right)
\exp\left(\ud\beta L_0\big[\phi_{k},\phi_{k+1}\big] \right).
\end{align}
In combination with the integral along the real-time as in \cref{eq:denominator}, the whole path integral is defined on a closed contour in the complex time plane, which is periodic along the imaginary time, with a period $\hbar \beta$.
Since there exist different ways to insert complete sets, there is some freedom in choosing the contour in the complex time plane.
For a graphic illustration, see \cref{fig:contour}.

So far, we have considered the density matrix of a general scalar field, but for free fields we can carry out the integrals in \cref{eq:density}.
It is more convenient to do this in momentum space, so that we introduce
\begin{align}
\phi(x) = \int \frac{\ud^d{ p}}{(2\pi)^d}\,\phi({ p})e^{i{ px}}.
\end{align}
Since $\phi$ is a real field, $\phi({ -p})=\phi({ p})^\dagger$, and we may write
\begin{align}
\phi({ p})=\phi_{\rm re}({ p})+i\phi_{\rm im}({ p}),
\end{align}
Thus it would be more appropriate to use its real and imaginary components as integration variables, in particular $\sqrt{2}\phi_{\rm re}(p)$ and $\sqrt{2}\phi_{\rm im}(p)$, which can be regarded as the result of a unitary transformation of $(\phi(p),\,\phi(-p))$.
On the other hand, one can also arrive at the same variables, by performing a real-to-real Fourier transform in the first place.
Later on, we will use $p,re,im$ to mean that it is these real integration variables that we use.
But it is easy to switch between $(\phi(p),\,\phi(-p))$ and $(\sqrt{2}\phi_{\rm re}(p),\,\sqrt{2}\phi_{\rm im}(p))$.
so that the free Lagrangian in momentum space takes the form,
\begin{align}
L_0\big[\phi_{k},\phi_{k+1}\big]=
\frac{1}{V}\sum_{p,re,im} \left[\frac{1}{2}\frac{\left(\phi_{k}(p)-\phi_{k+1}(p)\right)^2}{\left(-i\hbar \ud\beta\right)^2}-\frac{\omega_p^2}{2} \frac{(\phi_k(p))^2+(\phi_{k+1}(p))^2}{2}\right],
\end{align}
where $\omega_p=\sqrt{p^2+m^2}$, and $V$ is the spatial volume.
\footnote{We will allow ourselves to readily switch between continuum and discrete notation, treating $\int \frac{\ud^dp}{(2\pi)^d}$ and $\frac{1}{V}\sum_p$, as being interchangeable.
\label{note1}}
We can now switch \cref{eq:density} into momentum space, and carry out the integrals, 
\begin{align}
\label{eq:thermal}
 &\langle \phi_0;t_0 | 
e^{-\beta \hat{H}}
|\phi_N;t_0\rangle 
\\
=&
\left[\frac{1}{2\pi V\hbar^2\ud\beta }\right]^{\frac{N(N_s)^d}{2}}
\prod_{p,re,im}\int\prod_{k=I}^{N-1} {\mathcal D}\phi_k
\exp\left(\ud\beta L_0\big[\phi_{0},\phi_{I}\big] \right)
\exp\left(\ud\beta L_0\big[\phi_{k},\phi_{k+1}\big] \right)
\nonumber\\
=&\prod_{p,re,im} \left(\frac{\omega_p}{ 2\pi V \hbar \sinh(\hbar \omega_p \beta)}\right)^{1/2}
\exp\left(
- \frac{\omega_p\left[ \cosh(\hbar \omega_p\beta) \left(\phi^2_N(p)+\phi^2_0(p)\right)-2\phi_N(p)\phi_0(p)\right]}{2\hbar V\sinh(\hbar \omega_p\beta)}
\right)
,
\nonumber
\end{align}
where the overall constant on the second line is changed due to the Fourier transform, and to reach the last line we take the limit $\ud\beta\to 0$.
We are now able to calculate the partition function as, 
\begin{align}
\label{eq:partition}
Z= \prod_{p,re,im}\int \ud\phi(p)\langle \phi;t_0 | 
e^{-\beta \hat{H}}
|\phi;t_0\rangle 
=
\prod_{p}\frac{1}{2\sinh(\hbar \omega_p\beta/2)}
=
\prod_{p}\left(\sum_{n_p=0}^{\infty} e^{-\hbar \omega_p\beta(n_p+1/2)} \right)
.
\end{align}

\subsection{Initial density matrix  for vacuum and $n$-particle states}
\label{sec:vacuum}

Alternatively, we can also derive everything from the $n$-particle eigenstates.
The free theory is equivalent to a sum of independent harmonic oscillators with different $\omega_p$.
Therefore, one can derive $n$-particle eigenstates for the free field theory as one does in the harmonic oscillator.
We will skip the details of the derivation and only provide the final formulae.

In momentum space, the vacuum wave function is\footnote{
The wave function here is understood as a stationary wave function.
With the time-dependent phase term $e^{-i\omega_pt/2}$, the wave function is the ground-state solution of Schr\"odinger functional equation (\ref{eq:schrodinger}), and the energy of the ground state is $\hbar\omega_p/2$.
}
\begin{align}
\langle \phi | {\rm vac} \rangle 
=&\prod_{p,re,im}\left(\frac{\omega_p}{V\hbar \pi}\right)^{1/4}\exp\left( -\frac{\omega_p\phi^2(p)}{2V\hbar}\right)
\\
=&\left(\prod_{p}\left(\frac{\omega_p}{V\hbar \pi}\right)^{1/4}\right)\exp\left( -\frac{1}{\hbar}\int\frac{\ud^dp}{(2\pi)^d}\frac{\omega_p\phi^2(p)}{2}\right).
\nonumber
\end{align}
With it, we can write the density matrix of the vacuum  state as,
\begin{align}
\langle \phi_0;t_0 | {\rm vac} \rangle \langle  {\rm vac}| \phi_n;t_0 \rangle 
= 
\prod_{p,re,im}\left(\frac{\omega_p}{V\hbar \pi}\right)^{1/2}\exp\left( -\frac{\omega_p}{V\hbar}\frac{\phi_0(p)\phi_0(p)+\phi_n(p)\phi_n(p)}{2}\right).
\label{eq:den_vacuum}
\end{align}
The wave function of the $n$-particle state is
\begin{align}
\label{eq:n-par}
\langle \phi|n \rangle =
\prod_{p,re,im}\left(\frac{\omega_p}{V\hbar \pi}\right)^{1/4}
\frac{1}{\sqrt{2^{n_p}n_p!}}
h_{n_p}\left(\sqrt{\frac{\omega_p}{V\hbar}}\phi(p) \right)
\exp\left( -\frac{1}{2}\left(\sqrt{\frac{\omega_p}{V\hbar}}\phi(p) \right)^2\right)
,
\end{align}
where the Hermite polynomial $h_n(z)$ is defined as:
\begin{align}
h_n(z)=e^{z^2/2}\left(z-\frac{\ud}{\ud z}\right)^n e^{-z^2/2}.
\end{align}
We can now compute the density matrix of any pure state or mixed state, as long as it can be expanded with $n$-particle states.
For instance, it is straightforward to calculate the density matrix for the thermal states, up to the partition function $Z$, 
\begin{align}
 &\langle \phi_0;t_0 | 
e^{-\beta\hat{H}}
|\phi_N;t_0\rangle 
\\
=&
\prod_{p,re,im}\left(\frac{\omega_p}{V\hbar \pi}\right)^{1/2}
\sum_{n_p=0}^{+\infty}
\frac{1}{\sqrt{2^{n_p}n_p!}}
h_{n_p}\left(\sqrt{\frac{\omega_p}{V\hbar}}\phi_0(p) \right)
\exp\left( -\frac{1}{2}\left(\sqrt{\frac{\omega_p}{V\hbar}}\phi_0(p) \right)^2\right)
\nonumber \\
&\frac{1}{\sqrt{2^{n_p}n_p!}}
h_{n_p}\left(\sqrt{\frac{\omega_p}{V\hbar}}\phi_N(p) \right)
\exp\left( -\frac{1}{2}\left(\sqrt{\frac{\omega_p}{V\hbar}}\phi_N(p) \right)^2\right)
e^{-\hbar \omega_p\beta \left(n_p+\frac{1}{2}\right)}
\nonumber \\
=&
\prod_{p,re,im}\left(\frac{\omega_p}{ 2\pi V \hbar \sinh(\hbar \omega_p \beta)}\right)^{1/2}
\exp\left(-\frac{\omega_p}{V\hbar} \frac{\cosh(\hbar \omega_p\beta)(\phi^2_0(p)+\phi^2_N(p))-2\phi_0(p)\phi_N(p)}{2\sinh(\hbar \omega_p\beta)}\right)
\nonumber
,
\end{align}
where to get the final expression, we have used Mehler's formula
\begin{align}
\sum_{n=0}^{+\infty}\frac{\left(w/2\right)^n}{n!}h_n(x)h_n(y) \exp\left(-(x^2+y^2)/2\right)=\frac{1}{\sqrt{1-w^2}}\exp\left(\frac{4xyw-(1+w^2)(x^2+y^2)}{2(1-w^2)}\right)
.
\end{align}
This result agrees with what we derived in \cref{eq:thermal}.
It is useful to check the exact density matrix with the partition function (\ref{eq:partition}),
\begin{align}
 &\langle \phi_0;t_0 | 
e^{-\beta \hat{H}}/Z
|\phi_N;t_0\rangle 
\\
=&\prod_{p,re,im} \left(\frac{\omega_p}{ \pi V \hbar }\frac{\sinh(\hbar \omega_p \beta/2)}{\cosh(\hbar \omega_p \beta/2)}\right)^{1/2}
\exp\left(
- \frac{\omega_p\left[ \cosh(\hbar \omega_p\beta) \left(\phi^2_N(p)+\phi^2_0(p)\right)-2\phi_N(p)\phi_0(p)\right]}{2\hbar V\sinh(\hbar \omega_p\beta)}
\right)
.
\nonumber
\end{align}
In the limit $\beta\to +\infty$, it becomes to (\ref{eq:den_vacuum}).
The density matrix of the thermal state at zero temperature gives the density matrix of the vacuum.
So we are going to stick with the free thermal density matrix in the following sections, and treat the vacuum state as a special case.

\subsection{Path integral with a free initial density matrix}
\label{sec:free}

Given a free initial density matrix, the full path integral has the general form,
\begin{align}
\label{eq:full_pm}
Z=&\int {\mathcal D}\phi
\exp\left(-\frac{1}{\hbar}\int\frac{\ud^dp}{(2\pi)^d} \frac{\omega_p\left(\cosh(\hbar \omega_p\beta)\left[(\phi_0^+)^2+(\phi_0^-)^2\right]-2\phi_0^+\phi_0^-\right)}{2\sinh(\hbar \omega_p\beta)}+\frac{i}{\hbar}\int_{\mathcal C} \ud t L\right)
,
\end{align}
or, in the $\phi^{cl}$ and $\phi^q$ basis,
\begin{align}
\label{eq:full_cq}
Z=&\int {\mathcal D}\phi
\exp\left(-\frac{1}{\hbar}\int\frac{\ud^dp}{(2\pi)^d}\omega_p\left[
\frac{(\phi_0^{cl})^2}{2n_p+1}
+\frac{(\phi_0^{q})^2}{4}(2n_p+1)
\right] +\frac{i}{\hbar}\int_{\mathcal C} \ud t L\right)
,
\end{align}
with the occupation number given by
\begin{align}
n_p=\frac{1}{e^{\hbar \omega_p\beta}-1}
.
\end{align}
The initial density matrix in (\ref{eq:full_cq}) implies that the field $\phi_0^{cl}$ is drawn from a normal distribution with the variance proportional to $2n_p+1$, while $\phi_0^q$ comes from a normal distribution with variance proportional to $1/(2n_p+1)$.
We can get a better understanding of this observation by integrating out $\phi_0^{q}$, noting that $\phi_0^q$ also appears in the last term of \cref{eq:full_cq}.
However, by assuming the theory to be free at $t_0$, we will not encounter any higher order terms of $\phi_0^q$,
\begin{align}\label{eq:phi0qphi1c}
\frac{i}{\hbar}
\int_{\mathcal C}\ud t L
&=\left(\frac{i}{\hbar \ud t}\right)\int \frac{d^dp}{(2\pi)^d}
\Bigg[
\phi_0^{cl}\phi_0^{q}
-\phi_1^{cl}\phi_0^{q}
-\frac{\omega_p^2\ud t^2}{2}\phi_0^{cl}\phi_0^{q}
+\cdots
\Bigg],
\end{align}
and we see that $\phi_0^q$ interacts only with $\phi_0^{cl}$ and $\phi_1^{cl}$.
After the integrating out $\phi_0^q$ the path integral takes the form, 
{\small
\begin{align}
\label{eq:full_integrate}
&\int {\mathcal D}\phi
\exp\left(-\frac{1}{\hbar}\int\frac{\ud^dp}{(2\pi)^d}\left[
\frac{\omega_p(\phi_0^{cl}(p))^2}{2n_p+1}
+\frac{1}{\omega_p(2n_p+1)}\left(\frac{\phi_1^{cl}-\phi_0^{cl}\left(1-\omega_p^2\ud t^2/2\right)}{\ud t}\right)^2
\right] +\frac{i}{\hbar}\int_{\mathcal C} \ud t L'\right),
\end{align}}%
where $L'$ denotes $L$ with all $\phi_0^q$ related terms removed.
One now recognizes the new term in the square bracket above as just the time derivative of the scalar, but with finite $\ud t$, 
\begin{align}\label{eq:phiDotDiscrete}
\dot\phi_0^{cl}=\frac{\phi_1^{cl}-\phi_0^{cl}\left(1-\omega_p^2\ud t^2/2\right)}{\ud t},
\end{align}
and we now see that the density matrix gives Gaussian distributions to $\phi_0^{cl}$ and $\dot\phi_0^{cl}$ with variances given by,
\begin{align}
\label{eq:initial}
\langle \phi_0^{cl}(p) \Big(\phi_0^{cl}(p')\Big)^\dagger\rangle = \frac{\hbar}{\omega_p}\left(n_p+\frac{1}{2}\right)(2\pi)^d\delta^d(p-p'),
\nonumber \\
\langle \dot\phi_0^{cl}(p) \Big(\dot\phi_0^{cl}(p')\Big)^\dagger\rangle = \omega_p\hbar\left(n_p+\frac{1}{2}\right)(2\pi)^d\delta^d(p-p').
\end{align}
In \cref{sec:critical_points}, we mentioned that in the perturbation theory of $\phi^q$, the leading order theory has linear $\phi^q$ terms in the exponent, and therefore one can integrate $\phi^q$ out and obtain the classical equation of motion.
There is still, however, the initial density matrix left.
This means that the initialization of the classical theory should respect the distribution (\ref{eq:initial}).
In practice, we can generate ensembles of initializations of $\phi_0^{cl}$ and $\phi_1^{cl}$ according to (\ref{eq:phiDotDiscrete}) and (\ref{eq:initial}), and then use (\ref{eq:classic}) to find the full classical history. 
As we will show below, this classical history may then be used as the starting point for our Monte Carlo simulation of the path integral, although the Monte Carlo process essentially washes out the memory of the classical history (except $\phi_0^{cl}$ and $\phi_1^{cl}$, which are held fixed for a given Monte Carlo run.). 

In the full quantum field theory, we also want to separate the initial density matrix contribution from the rest of the closed time path in the path integral.
There are two reasons for doing this: 

(1) It is much easier to write the initial density matrix part in momentum space, and the subsequent dynamical part of the path integral in configuration space.

(2) There is no ``sign problem'' in the initial density matrix piece.

\noindent
In fact, the distributions in the initial density matrix piece of (\ref{eq:initial}) are ordinary Gaussian distributions, and simple Monte Carlo methods are sufficient to generate samples of  $\phi_0^{cl}$ and $\phi_1^{cl}$.
Thus, in addition, we also want to treat $\phi_0^{cl}$ and $\phi_1^{cl}$ on a different footing from the other integration variables.
However, while the initial density matrix part involves only $\phi_0^{cl}$ and $\phi_1^{cl}$, the remaining part of the path integral also contains $\phi_0^{cl}$ and $\phi_1^{cl}$.
So is the separation legitimate?
The answer is yes, but with a note of caution.

\subsection{Separating variables}
\label{sec:separating}

When separating $\phi_0^{cl}$ and $\phi_1^{cl}$ from the other integration variables, we should check that the following equality is valid,
\begin{align}
\label{eq:separate}
&\frac{\int {\mathcal D}\phi_0^{cl} {\mathcal D}\phi_1^{cl}
\rho\left(\phi_0^{cl},\phi_1^{cl}\right)
\int \prod_{{i}=1}^{m-1}{\mathcal D}\phi_{{i}}^{q}{\mathcal D}\phi_{{i}+1}^{cl} 
\exp\left( \frac{i}{\hbar} \int_{\mathcal C}\ud tL'\right)
\mathcal{O}}{
\int {\mathcal D}\phi_0^{cl} {\mathcal D}\phi_1^{cl}
\rho\left(\phi_0^{cl},\phi_1^{cl}\right)
\int \prod_{{i}=1}^{m-1}{\mathcal D}\phi_{{i}}^{q}{\mathcal D}\phi_{{i}+1}^{cl} 
\exp\left( \frac{i}{\hbar} \int_{\mathcal C}\ud tL'\right)
}
\nonumber \\
=&
\frac{\int {\mathcal D}\phi_0^{cl} {\mathcal D}\phi_1^{cl}
\rho\left(\phi_0^{cl},\phi_1^{cl}\right)
\Bigg[\frac{
\int \prod_{{i}=1}^{m-1}{\mathcal D}\phi_{{i}}^{q}{\mathcal D}\phi_{{i}+1}^{cl} 
\exp\left( \frac{i}{\hbar} \int_{\mathcal C}\ud tL'\right)
\mathcal{O}}{
\int \prod_{{i}=1}^{m-1}{\mathcal D}\phi_{{i}}^{q}{\mathcal D}\phi_{{i}+1}^{cl} 
\exp\left( \frac{i}{\hbar} \int_{\mathcal C}\ud tL'\right)
}\Bigg]
}{
\int {\mathcal D}\phi_0^{cl} {\mathcal D}\phi_1^{cl}
\rho\left(\phi_0^{cl},\phi_1^{cl}\right)
},
\end{align}
where $\rho\left(\phi_0^{cl},\phi_1^{cl}\right)$ is the density matrix part in \cref{eq:full_integrate}, and is a function of $\phi_0^{cl}$ and $\phi_1^{cl}$ only.
Apparently, to have the equality valid, the lifted integral should be independent of $\phi_0^{cl}$ and $\phi_1^{cl}$.
To show that this is true, we make use of a feature that we have already explored: 
The only term in $L'$ containing $\phi_{m}^{cl}$ is from $\phi_m^{cl}(x)\phi_{m-1}^q(x)$, and by integrating out $\phi_m^{cl}$, we obtain a delta function, $\delta(\phi_{m-1}^{q})$. Then by integrating out $\phi_{m-1}^q$, we obtain an integral similar to the previous one, but with $\phi_{m-1}^{cl}$ now playing the role of $\phi_{m}^{cl}$.
We can continue this contraction of the closed time path down to $\phi_1^{q}$, where we then find $\delta(\phi_1^{q})$.
Now, we know that all $\phi_0^{cl}$ and $\phi_1^{cl}$ appear in $L'$ only through their products with $\phi_1^{q}$, so
by integrating out the delta function of $\phi_1^q$, we know the result has no dependence on $\phi_0^{cl}$ and $\phi_1^{cl}$.
Concretely, the result of the integral is 
\begin{align}
\int \prod_{{i}=1}^{m-1}{\mathcal D}\phi_{{i}}^{q}{\mathcal D}\phi_{{i}+1}^{cl} 
\exp\left( \frac{i}{\hbar} \int_{\mathcal C}\ud tL'\right)
=\left(\frac{2\pi\hbar \ud t}{\ud^dx}\right)^{(N_s)^d(m-1)}.
\label{eq:constant}
\end{align}
which is independent of $\phi_0^{cl}$ and $\phi_1^{cl}$, and so a constant from the point of view of the integral over initial conditions. We may thus perform the separation of variables in (\ref{eq:separate}).

\subsection{One critical point for one initialization}
\label{sec:one}

We separate the whole path integral into two parts: the initial density matrix and the rest of the path integral. To implement the Monte Carlo simulation, we propose different algorithms for each of these different parts.

1. We assume the initial density matrix is known, so we  can sample $\phi_0^{cl}$ and $\phi_1^{cl}$ directly according to the initial density matrix, using simple Monte Carlo algorithms.
There is no ``sign problem'' in the procedure, as in  momentum space the distribution function is real and vanishes exponentially as $|\phi|\to\infty$ \cite{Bender:1969si}.
Notice that the initial density matrix is a function of $\phi_0^{cl}$ and $\phi_1^{cl}$ only, but the rest of the path integral also depends on $\phi_0^{cl}$ and $\phi_1^{cl}$.
We denote such sampled fields as $\tilde \phi_0^{cl}$ and $\tilde \phi_1^{cl}$, and a Fourier transform is necessary to bring the fields into configuration space for later use.
All these $\tilde \phi_0^{cl}(x)$ and $\tilde \phi_1^{cl}(x)$ are real.

2. Provided with each $\tilde \phi_0^{cl}$ and $\tilde \phi_1^{cl}$, we then perform importance sampling according to 
\begin{align}
\label{eq:single}
\int \prod_{{i}=1}^{m-1}{\mathcal D}\phi_{{i}}^{q}{\mathcal D}\phi_{{i}+1}^{cl} 
\exp\left( \frac{i}{\hbar} \int_{\mathcal C}\ud tL'\right)
,
\end{align}
with the Generalized Thimble Method, according to an algorithm such as in \cite{Alexandru:2017lqr}. Note that the quantum and classical fields start at 1 and 2 in the product, respectively, because $\phi_{0}^{q}$ has been integrated out, while $\phi_{1}^{cl}$ and $\phi_{2}^{cl}$ are specified as initial data for each initialization. 
The sampled $\phi_{{i}+1}^{cl}$ and $\phi_{{i}}^{q}$ with $1\leq {i}\leq m-1$ in this procedure are complex.
With reweighting (\ref{eq:reweighting}), we can calculate the expectation value of an operator $\hat{\mathcal O}$ over a single initialization, which is equivalent to,
\begin{align}
\langle \hat{\mathcal O}\rangle_{single} =
\frac{
\int \prod_{{i}=1}^{m-1}{\mathcal D}\phi_{{i}}^{q}{\mathcal D}\phi_{{i}+1}^{cl} 
\exp\left( \frac{i}{\hbar} \int_{\mathcal C}\ud tL'\right)
\mathcal{O}}{
\int \prod_{{i}=1}^{m-1}{\mathcal D}\phi_{{i}}^{q}{\mathcal D}\phi_{{i}+1}^{cl} 
\exp\left( \frac{i}{\hbar} \int_{\mathcal C}\ud tL'\right)
}.
\end{align}
The full expectation, $\langle \hat{\mathcal O}\rangle$, in \cref{eq:separate} will then be the mean of all the singles, $\langle \hat{\mathcal O}\rangle_{single}$.

For the integral (\ref{eq:single}) above, we can repeat the analysis in \cref{sec:critical_points} to find all the critical points, this time with ${\mathcal I}=-i\int_{\mathcal C}\ud t L'/\hbar$.
In fact, the conclusions in \cref{sec:critical_points} are still valid here:
At critical points, all $\phi_{{i}}^q(x)=0$, so ${\mathcal I=0}$, as it consists of odd terms of $\phi^q$, and all $\phi_{{i}+1}^{cl}(x)$ are uniquely determined through the classical equation of motion (\ref{eq:classic}), once $\tilde \phi_0^{cl}$ and $\tilde \phi_1^{cl}$ are specified.
In other words, for each initialization, there exists one and only one critical point.
This means that for step 2 above,  we will not encounter any multimodal problem that would be caused by the existence of multiple critical points.

However, the initial density matrix could possess multiple saddle points in its distribution.
For instance, we expect this to happen in the density matrix of $n$-particle state when $n\neq 0$, or in the case of multi-scalar fields where there exists some symmetry among those scalars.
Still, this will not change the conclusion that there exists one and only one critical point for the thimble part of the calculation, and we only need to deal with one thimble/critical point on step 2.

We stress that the derivation is valid on the complexified fields, and the thimble must contribute to the original integral, as the critical point is located on the real field plane.
There is one more thing we can predict.
With each initialization, the averaged phase $\langle e^{-i{\rm Im}[{\mathcal I}]+i{\rm arg}\left({\rm det}(J)\right)} \rangle_P $ must be real and positive, due to \cref{eq:constant}.
Furthermore, on the Lefschetz thimble, ${\mathcal I}$ vanishes at the critical point, so ${\rm Im}[{\mathcal I}] =0$ on the whole thimble, and only the residual phase ${\rm arg}\left({\rm det}(J)\right)$ contributes.

\subsection{Two-point functions}
\label{sec:correlator}

In order to test the formalism we will calculate the two-point correlators analytically, and compare them with numerical results based on the procedure described above.
One can do this in the framework of perturbation theory, that is we first compute free correlators and then add the loop corrections.
In this section, we only explicitly derive the free two-point functions, while a 1-loop correction will be included in App \ref{app:loop}. See also \cite{Aarts:1997kp}.
Since in the free theory, different momentum modes are independent of each other, we can focus the calculation on a single mode.
There are two equivalent ways, up to a constant due to the integration of $\phi_0^q$, to write the path integral,
\begin{align}
Z=&\quad\quad \quad \int {\mathcal D}\phi^{+} {\mathcal D}\phi^{-}
\exp\left(-\frac{\omega_p}{V\hbar} \frac{\cosh(\hbar \omega_p\beta)\left[(\phi_0^+)^2+(\phi_0^-)^2\right]-2\phi_0^+\phi_0^-}{2\sinh(\hbar \omega_p\beta)}\right)
\nonumber\\
\label{eq:exp1}
&\exp\left(\left(\frac{i\ud t}{V\hbar}\right) \sum_{{i}=0}^{m-1}
\left[\frac{1}{2}\left(\frac{\phi_{{i}+1}^{+}-\phi_{{i}}^{+}}{\ud t}\right)^2
-\frac{\omega_p^2}{2}\frac{\left(\phi_{{i}+1}^{+}\right)^2+\left(\phi_{{i}}^{+}\right)^2}{2}
-\Big(\phi^+\to\phi^-\Big)
\right]
\right),
\end{align}
\begin{align}
Z=& \int {\mathcal D}\phi_0^{cl} {\mathcal D}\phi_1^{cl}
\exp\left(-\frac{1}{V\hbar} 
\left[
\frac{\omega_p}{2n_p+1}(\phi_0^{cl}(p))^2
+\frac{1}{\omega_p(2n_p+1)}\left(\frac{\phi_1^{cl}-\phi_0^{cl}\cos(\tilde\omega_p\ud t)}{\ud t}\right)^2
\right]
\right)
\nonumber \\
\label{eq:exp2}
&\int \prod_{{i}=1}^{m-1}{\mathcal D}\phi_{{i}}^{q}{\mathcal D}\phi_{{i}+1}^{cl} 
\exp\left( 
\left(\frac{i}{V\hbar \ud t}\right)
\phi_{{i}}^{q}(p)\left[2\cos(\tilde\omega_p\ud t)\phi_{{i}}^{cl}(p)-\phi_{{i}-1}^{cl}(p)-\phi_{{i}+1}^{cl}(p)\right]
\right),
\end{align}
with constants
\begin{align}
\label{eq:omega_tilde_def}
n_p=\frac{1}{e^{\hbar \omega_p\beta}-1}
,\quad
\cos(\tilde\omega_p\ud t)\stackrel{!}{=} 1-\frac{\omega_p^2\ud t^2}{2},
\end{align}
where $\omega_p$ is the frequency in the continuous theory but, because of the discretization, it is $\tilde\omega_p$ that propagates on the lattice.
In the limit $\ud t\to 0$, $\tilde\omega_p$ converges to $\omega_p$.
For finite $\ud t$, it is convenient to replace $\omega_p$ in (\ref{eq:exp1}) and (\ref{eq:exp2}) with $\sin(\tilde\omega_p\ud t)/\ud t$.
With only Gaussian functions in (\ref{eq:exp1}) and (\ref{eq:exp2}), we can calculate the free two-point functions as,
\begin{align}
\langle xx^T\rangle_0=\frac{\int d^nx~xx^Te^{-x^TAx}}{\int d^nx~e^{-x^TAx}}=\frac{A^{-1}}{2},
\label{eq:toy}
\end{align}
where $A$ and $x$ are understood to be a symmetric complex matrix and a real vector respectively. The size is given by the number of discrete points on the time contour of choice.
The above normalization is appropriate for the discrete theory, while for the continuous theory, there will exist a factor of $V$ in the definition.
To compensate this, we simply assume $V=1$ in the following derivation.

\subsection{Time-ordered correlators }
\label{subsec:feynman}

It is straightforward to identify the matrix $A$ in \cref{eq:exp1}, then calculate its inverse, and use (\ref{eq:toy}) to discover that the two-point functions in the $(\phi^+,~\phi^-)$ basis are 
{\small
\begin{align}
\label{eq:+-}
\left(
\begin{array}{ccccccc}
 \langle\phi_0^{+}\phi_0^{+}\rangle_0 & \langle\phi_0^{+}\phi_1^{+}\rangle_0 &\cdots  & \langle\phi_0^{+}\phi_m\rangle_0 & \cdots & \langle\phi_0^{+}\phi_1^-\rangle_0 & \langle\phi_0^{+}\phi_0^-\rangle_0  \\
 \langle\phi_1^{+}\phi_0^{+}\rangle_0 & \langle\phi_1^{+}\phi_1^{+}\rangle_0 &\cdots  & \langle\phi_1^{+}\phi_m\rangle_0 & \cdots & \langle\phi_1^{+}\phi_1^-\rangle_0 & \langle\phi_1^{+}\phi_0^-\rangle_0  \\
\vdots &\vdots &\ddots &\vdots &\iddots &\vdots &\vdots\\
 \langle\phi_m\phi_0^{+}\rangle_0 & \langle\phi_m\phi_1^{+}\rangle_0 &\cdots  & \langle\phi_m\phi_m\rangle_0 & \cdots & \langle\phi_m\phi_1^-\rangle_0 & \langle\phi_m\phi_0^-\rangle_0  \\
\vdots &\vdots &\iddots &\vdots &\ddots &\vdots &\vdots\\
 \langle\phi_1^-\phi_0^{+}\rangle_0 & \langle\phi_1^-\phi_1^{+}\rangle_0 &\cdots  & \langle\phi_1^-\phi_m\rangle_0 & \cdots & \langle\phi_1^-\phi_1^-\rangle_0 & \langle\phi_1^-\phi_0^-\rangle_0  \\
 \langle\phi_0^-\phi_0^{+}\rangle_0 & \langle\phi_0^-\phi_1^{+}\rangle_0 &\cdots  & \langle\phi_0^-\phi_m\rangle_0 & \cdots & \langle\phi_0^-\phi_1^-\rangle_0 & \langle\phi_0^-\phi_0^-\rangle_0  
\end{array}
\right)=
\frac{\hbar \ud t}{\sin(\tilde\omega_p\ud t)} 
\left( \frac{n_p+1}{2}F+\frac{n_p}{2}F^*\right),
\end{align}
}%
where the star denotes complex conjugation, and  the matrix $F$ is 
{\small
\begin{align}
&F=
\left(
\begin{array}{ccccccc}
 1 &  e^{-i\tilde\omega_p\ud t} &  \cdots & e^{-im\tilde\omega_p\ud t} &  \cdots & e^{-i\tilde\omega_p\ud t} & 1 \\
e^{-i\tilde\omega_p\ud t} & 1 & \cdots & e^{-i[m-1]\tilde\omega_p\ud t} & \cdots & 1 & e^{i\tilde\omega_p\ud t} \\
\vdots &\vdots &\ddots &\vdots &\iddots &\vdots &\vdots\\
e^{-im\tilde\omega_p\ud t} & e^{-i[m-1]\tilde\omega_p\ud t}& \cdots & 1 & \cdots  &  e^{i[m-1]\tilde\omega_p\ud t}& e^{im\tilde\omega_p\ud t} \\
\vdots &\vdots &\iddots &\vdots &\ddots &\vdots &\vdots\\
e^{-i\tilde\omega_p\ud t} & 1 & \cdots & e^{i[m-1]\tilde\omega_p\ud t} & \cdots & 1 & e^{i\tilde\omega_p\ud t} \\
 1 &  e^{i\tilde\omega_p\ud t} &  \cdots & e^{im\tilde\omega_p\ud t} &  \cdots & e^{i\tilde\omega_p\ud t} & 1 
\end{array}
\right).
\end{align}
}%
There are two features worth emphasizing in the above expression.\\

1. 
In the vacuum, that is $n_p=0$, we notice that the rows and columns corresponding to $\phi_0^+\to\phi_m$ (i.e. the upper-left part of $F$) lead to $F_{jk}=\exp\left( -i\omega_p|t_j-t_k| \right)$, and give the Feynman propagator, which is defined as
\footnote{To obtain the Feynman propagator in $d+1$ dimension, one can first do the Fourier transform to get the two-point function in the momentum space.
Since two-point correlators with different frequencies vanish, one can then write the final expression as a sum or integral over momentum, where we presume the sum and integral to be interchangeable, see also \cref{note1}.
}
\begin{align}\label{eq:feyn_prop}
&-i\langle 0 | T \Phi(x) \Phi(y) | 0\rangle _0
=\hbar\int \frac{d\omega}{2\pi}\frac{d^dp}{(2\pi)^d} \frac{e^{-i\omega(t_x-t_y)+ip(x-y)}}{\omega^2-p^2-m^2+i\epsilon}
=-i\hbar\int\frac{d^dp}{(2\pi)^d} \frac{e^{-i\omega_p|t_x-t_y|+ip(x-y)}}{2\omega_p}.
\end{align}
Thus  we get the correct $i\epsilon$ prescription in the propagator.
This also means the correlators $\langle \phi_i^+ \phi_j^+\rangle_0$ are time-ordered, while the correlators $\langle \phi_i^- \phi_j^-\rangle_0$ are anti-time-ordered. 
On the other hand, when $n_p\neq 0$, we can calculate the equal-time correlator through summing the Matsubara frequencies, 
\begin{align}
&\langle 0 | \Phi(x) \Phi(y) | 0\rangle 
=-\frac{\hbar}{\hbar \beta}\sum_n\int \frac{d^dp}{(2\pi)^d} \frac{e^{ip(x-y)}}{(i2\pi n/(\hbar \beta))^2-\omega_p^2}
=\hbar\int\frac{d^dp}{(2\pi)^d} e^{ip(x-y)}\frac{2n_p+1}{2\omega_p}.
\end{align}
This corresponds to calculating the equal-time elements in \cref{eq:+-}.
\\

2. There exist symmetries in the above two-point functions. 
For instance, $\langle \phi_i^+ \phi_j^+\rangle_0$ =$\langle \phi_i^- \phi_j^+\rangle_0 $ if $i>j$.
In fact, although we can have many integration variables $\phi_i$ at time $t_i$, there is only one operator $\hat\Phi_i$,
and it is actually easier to discern the symmetries from the operator formalism,
\begin{align}
\label{eq:G}
&\langle \phi_i^+ \phi_j^+\rangle =\theta(t_i-t_j)G^{>}+\theta(t_j-t_i)G^{<},
\nonumber \\
&\langle \phi_i^+ \phi_j^-\rangle=G^{<},\qquad \qquad \langle \phi_i^- \phi_j^+\rangle=G^{>},
\nonumber \\
&\langle \phi_i^- \phi_j^-\rangle=\theta(t_j-t_i)G^{>}+\theta(t_i-t_j)G^{<},
\end{align}
with 
\begin{align}
G^{>}=\langle \hat\Phi_i\hat\Phi_j\rangle,\qquad
G^{<}=\langle \hat\Phi_j\hat\Phi_i\rangle.
\end{align}
On the other hand, as in \cref{eq:+-}, 
\begin{align}
G_0^{>}(t_i-t_j)&\propto e^{-i(t_i-t_j)\tilde\omega_p}(n_p+1)+e^{i(t_i-t_j)\tilde\omega_p}n_p
,\nonumber \\
G_0^{<}(t_i-t_j)&\propto e^{i(t_i-t_j)\tilde\omega_p}(n_p+1)+e^{-i(t_i-t_j)\tilde\omega_p}n_p,
\end{align}
and this makes manifest the KMS condition $G^>(t_i-t_j)=G^{<}(t_i-t_j+i\hbar\beta)$ \cite{Aarts:1997kp}.

\subsection{Classical-Classical and Quantum-Classical correlators}
\label{sec:CQQQCorr}

We could obtain the correlators such as $\phi_i^{cl}\phi_j^{cl}$ or $\phi_i^{q}\phi_j^{cl}$ through a rotation of $\phi_i^{\pm}\phi_j^{\pm}$ in \cref{eq:+-},
but it is instructive to derive the expression from scratch with a simple example.
Consider $m=3$.
Then the matrix $A$ in \cref{eq:exp2} is
{\small
\begin{align}
\label{eq:A}
A=\left(
\begin{array}{cccc:cc}
a & -a\cos(\tilde\omega_p\ud t) & 0  & 0  & -b&0\\
-a\cos(\tilde\omega_p\ud t) & a & 0 & 0 & 2b\cos(\tilde\omega_p\ud t) & -b \\
0 & 0 & 0  & 0 & -b & 2b\cos(\tilde\omega_p\ud t)\\
0 & 0 & 0  & 0&0 &-b\\
\hdashline 
-b & 2b\cos(\tilde\omega_p\ud t)& -b & 0 & 0 & 0\\
0 & -b & 2b\cos(\tilde\omega_p\ud t)& -b & 0 & 0 
\end{array}
\right)
,~~
x=
\left(
\begin{array}{c}
\phi_{0}^{cl} \\
\phi_{1}^{cl} \\
\phi_{2}^{cl} \\
\phi_{3} \\
\phi_{1}^{q} \\
\phi_{2}^{q} 
\end{array}
\right)
,
\end{align}
}%
with constants
\begin{align}
a=\frac{1}{\hbar (2n_p+1)\ud t\sin(\tilde\omega_p\ud t)}
,\quad
b=-\frac{i}{2\ud t\hbar}.
\end{align}
We treat $\phi_m$ as a $\phi^{cl}$ field.
Since we have also integrated out $\phi_0^{q}$, in the end there are two more  $\phi^{cl}$ fields than $\phi^q$ fields.
Following \cref{eq:toy}, we arrive at 
{\small
\begin{align}
\left(
\begin{array}{ccc:c}
& & & \\
& \langle\phi^{cl}\phi^{cl}\rangle &  &\langle\phi^{cl}\phi^{q}\rangle \\
& & & \\
\hdashline & & & \\ [-0.4cm] 
& \langle\phi^{q}\phi^{cl}\rangle &  &\langle\phi^{q}\phi^{q}\rangle\\
\end{array}
\right)
=
\left(
\begin{array}{cccc:cc}
f & f\cos(\tilde\omega_p\ud t) &  f\cos(2\tilde\omega_p\ud t) &  f\cos(3\tilde\omega_p\ud t) & 0 & 0 \\
f\cos(\tilde\omega_p\ud t) & f &  f\cos(\tilde\omega_p\ud t)  &  f\cos(2\tilde\omega_p\ud t)  & 0 & 0 \\
f\cos(2\tilde\omega_p\ud t) & f\cos(\tilde\omega_p\ud t)  & f &  f\cos(\tilde\omega_p\ud t)  & r\sin(\tilde\omega_p\ud t) & 0 \\
f\cos(3\tilde\omega_p\ud t) & f\cos(2\tilde\omega_p\ud t) &  f\cos(\tilde\omega_p\ud t)  & f  & r\sin(2\tilde\omega_p\ud t) & r\sin(\tilde\omega_p\ud t)  \\
\hdashline 
 0 & 0 & r\sin(\tilde\omega_p\ud t) &  r\sin(2\tilde\omega_p\ud t) & 0 & 0 \\
 0 & 0 & 0 &  r\sin(\tilde\omega_p\ud t) & 0 & 0 
\end{array}
\right)
,
\end{align}}%
where
\begin{align}
f=\left(n_p+\frac{1}{2}\right)\frac{\hbar \ud t}{\sin(\tilde\omega_p\ud t)}
,\quad
r=-\frac{i\hbar \ud t}{\sin(\tilde\omega_p\ud t)}.
\end{align}
This may be summarized by the following:
\begin{align}
\label{eq:cl-cl}
\langle\phi^{cl}_i\phi^{cl}_j\rangle_0&=\hbar\left(n_p+\frac{1}{2}\right)\frac{\ud t}{\sin(\tilde\omega_p\ud t)}\cos(\tilde\omega_p(i-j)\ud t),\\
\label{eq:cl-q}
\langle\phi^{cl}_i\phi^{q}_j\rangle_0&=-i\hbar\theta(i-j)\frac{\ud t}{\sin(\tilde\omega_p\ud t)}\sin(\tilde\omega_p(i-j)\ud t),\\
\label{eq:q-cl}
\langle\phi^{q}_i\phi^{cl}_j\rangle_0&=-i\hbar\theta(j-i)\frac{\ud t}{\sin(\tilde\omega_p\ud t)}\sin(\tilde\omega_p(j-i)\ud t),\\
\label{eq:q-q}
\langle\phi^{q}_i\phi^{q}_j\rangle_0&=0,\\
\theta(i-j)&=\left\{
\begin{array}{cc}
1	&	\qquad i>j,\\
0	&	\qquad i\leq j.
\end{array}
\right.
\end{align}
We see, for example, that the correlators  $\langle\phi_i^{q}\phi_j^{cl}\rangle$ vanish unless $i<j$, and so correspond to the advanced propagators.
Furthermore, because of the advanced propagators,  any loop correction will not alter $\langle\phi^{q}\phi^{q}\rangle=0$.
Actually, we can derive this conclusion much more quickly from the operator formalism (\ref{eq:G}): $\langle\phi^{q}\phi^{q}\rangle=\langle\phi^{+}\phi^{+}\rangle+\langle\phi^{-}\phi^{-}\rangle-\langle\phi^{+}\phi^{-}\rangle-\langle\phi^{-}\phi^{+}\rangle=0$.

\section{Numerical Simulation}
\label{sec:numerics}

We now demonstrate how to carry out numerical simulations, with an example of $\lambda \phi^4$ theory (see also \cite{Alexandru:2017lqr,Aarts:1997kp}), using the following action,
\begin{align}
S=\int \ud t\ud^dx\left[\frac{1}{2}\dot\phi^2-\frac{1}{2}\left(\nabla\phi\right)^2-\frac{1}{2}m^2\phi^2-\frac{\lambda}{4!}\phi^4\right].
\end{align}
Ideally, we would like to simulate a $1+1$ or even $3+1$-dimensional system.
But in those cases, one should stick with some specific renormalization scheme in order to compare with the result of continuum theory.
This is beyond the scope of the present work, and is postponed for later work.
Instead, we find it is straightforward to compare with theoretical predictions in $0+1$-dimensional system, so quantum mechanics\footnote{For the application of Lefschetz thimble on quantum mechanics from a different perspective, see \cite{Tanizaki:2014xba,Cherman:2014sba}.}, where no divergence exists, and therefore no renormalization scheme is required. 
We shall set up our definitions in $d=1$ spatial dimensions, whereas in the actual simulations presented here, we have further reduced to $d=0$ quantum mechanics.  
Throughout the paper, we set $mdt=0.75$ for small couplings, and $mdt=0.5$ for large couplings, (more details in our future publications).

Space is discretized on $N_s$ sites, with periodic boundary conditions, and the time direction is discretized as above onto $N_t=2m+1$ sites going back and forth on the Keldysh contour (see \cref{fig:seq}). 

\subsection{Warm-up: Classical statistical approximation}
\label{sec:classtat}

We set the initial $\phi_0^{cl}(p)$ and $\phi_1^{cl}(p)$ according to \cref{eq:initial}, a Gaussian thermal density matrix.{\footnote{
For initial $n$-particle states, one could use the expression given in \cref{eq:n-par}, with some Hermite polynomial function.}
Given the distribution, we generate random samples of momentum-space variables $\phi_0^{cl}(p)$ and $\phi_1^{cl}(p)$ which are then Fourier transformed to position space $\phi_0^{cl}(x)$ and $\phi_1^{cl}(x)$. 
Now, we can compute the classical field evolution through the equation of motion,
{\small
\begin{align}
\label{eq:eom}
\frac{\tilde\phi_{i+1}^{cl}(x) -2\tilde\phi_{i}^{cl}(x)+\tilde\phi_{i-1}^{cl}(x)}{dt^2}
-
\frac{\tilde\phi_{i}^{cl}(x+1) -2\tilde\phi_{i}^{cl}(x)+\tilde\phi_{i}^{cl}(x-1)}{dx^2}
+
m^2\tilde\phi_{i}^{cl}(x)+\frac{\lambda}{6}\left(\tilde\phi_{i}^{cl}(x)\right)^3
=0.
\end{align}
}We use $\tilde\phi$ to refer to the fact that these are not variables of integration in the path integral. They represent the critical configuration in our complexified field configuration space, $\phi^{cl}=\tilde\phi^{cl},~\phi^q=0$, from which we will initiate our Monte-Carlo simulation in later sections. 

\begin{figure}[t]
\begin{tabular}{lr}
\hspace{-1.2cm}
\includegraphics[width=0.6\textwidth]{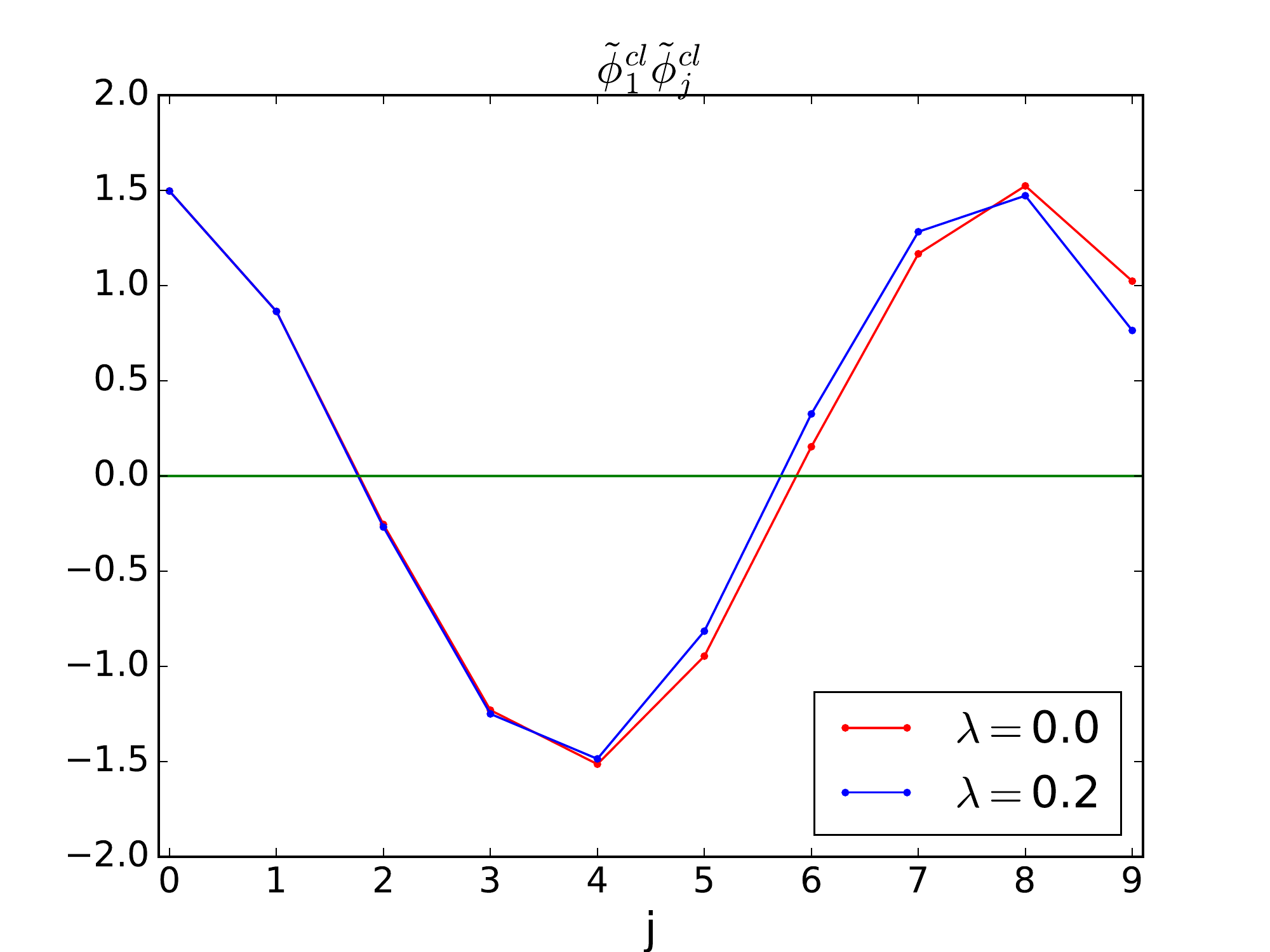} & \hspace{-1.2cm}
\includegraphics[width=0.6\textwidth]{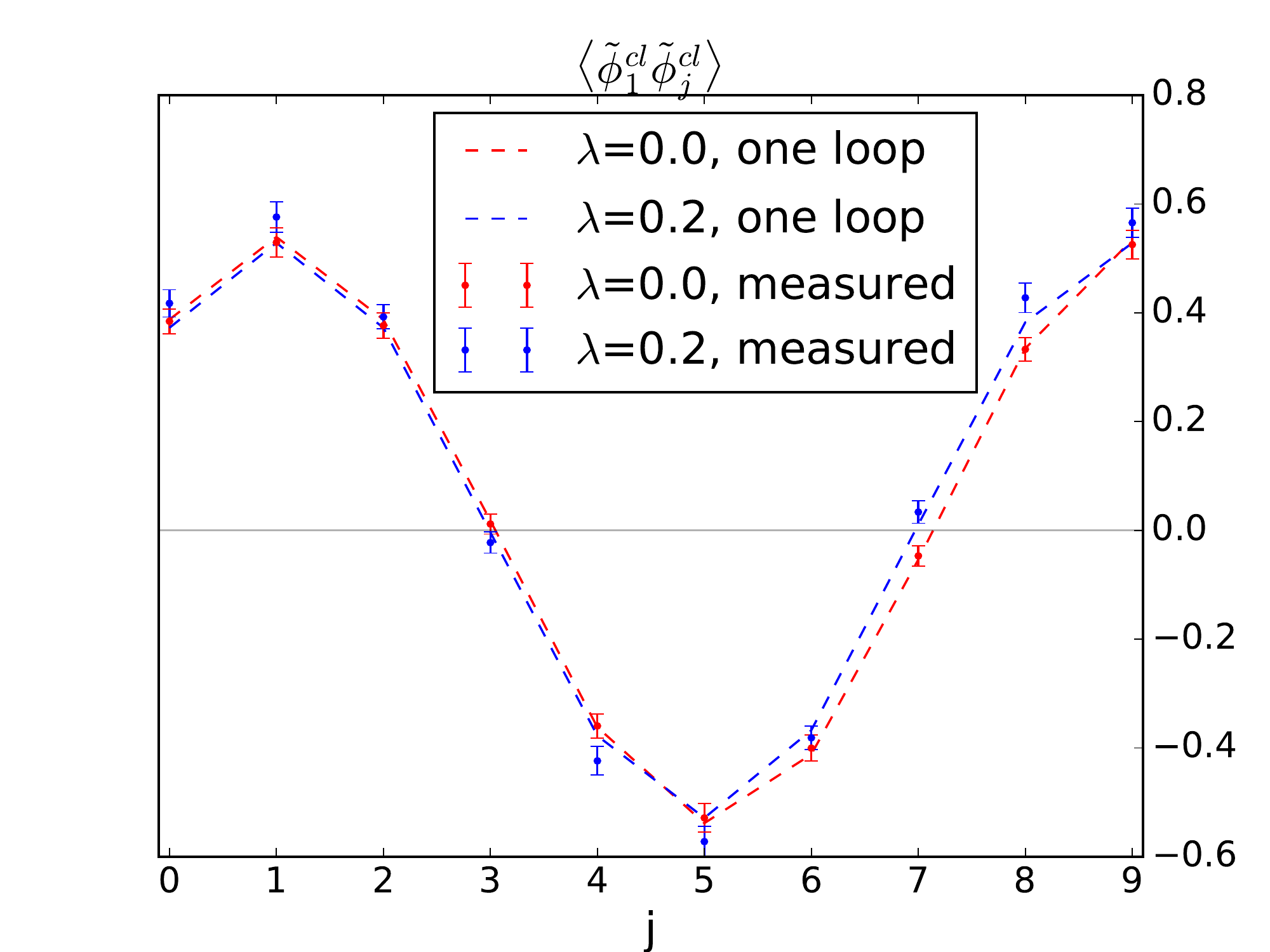}
\end{tabular}
\caption{Correlators for a single classical realisation (left) and averaged over initial conditions (right).}
\label{fig:classim}
\end{figure}

\Cref{fig:classim} (left) shows the correlator for a single such classical trajectory. In a classical simulation, we can only compute the classical-classical correlator. By averaging over the ensemble of initial conditions, we recover the ``classical-statistical" approximation to quantum dynamics, shown in \cref{fig:classim} (right). We show the results for a free field, $\lambda=0$ and an interacting theory $\lambda=0.2$. The correlators are very similar, but deviate enough that we can tell the difference with moderate statistics. The loop calculation is discussed in appendix \ref{app:loop}, where it is found that at 1-loop we just need to make the replacement $\omega_p^2\to \omega_p^2+\frac{\hbar\lambda}{4\omega}$. This is substituted into (\ref{eq:omega_tilde_def}) to find $\tilde\omega_p$, which is then used in expression (\ref{eq:cl-cl}) for the classical-classical correlator.

\subsection{Warm-up: Quantum average of a single initial realisation}
\label{sec:quantumaverage}

\begin{figure}[t]
\centering
\includegraphics[width=0.65\textwidth]{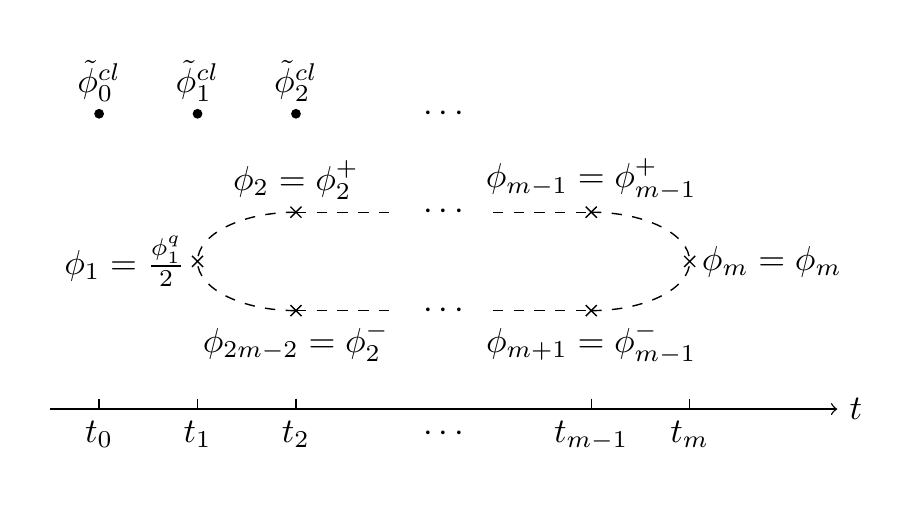}
\caption{The variables to be integrated over on the real-time contour, after the initial conditions are fixed.}
\label{fig:denotion}
\end{figure}

Going beyond the classical approximation then amounts to performing the complete path integral, the integrations of all the field variables not associated with the initial conditon, see \cref{fig:denotion}. 
As in \cref{sec:one}, we can write the integrand as $e^{-\mathcal I}$, with ${\mathcal I}=-i\int_{\mathcal C}\ud t L'/\hbar$.
It turns out that the exponent ${\mathcal I}$ is more conveniently expressed in the $(\phi^+,~\phi^-)$ basis than using $(\phi^{cl},~\phi^q)$, as the interaction terms are simpler there. We therefore switch to $(\phi^+,~\phi^-)$, except that at $t_1$ should be treated differently, since we count $\phi_1^{cl}$ into the initial condition, leaving $\phi_1^q$ as the only variable at $t_1$.
The exponent ${\mathcal I}$ also contains $\tilde\phi_0^{cl}$ and $\tilde\phi_1^{cl}$, and may be written as
{\small
\begin{align}
\label{eq:I}
&{\mathcal I}=\left(\frac{-i\ud x}{\hbar }\right)\sum_x\Bigg\{
2\phi_1(x)\frac{\tilde\phi_2^{cl}(x)}{\ud t}
-\frac{\lambda \ud t}{3}\tilde\phi_1^{cl}(x)\big(\phi_1(x)\big)^3
-\phi_{2}(x)\frac{\tilde\phi_1^{cl}(x)}{\ud t}
+\phi_{2m-2}(x)\frac{\tilde\phi_1^{cl}(x)}{\ud t}+
\nonumber \\
&\sum_{i=1}^{2m-2}
\frac{\big[\phi_{i+1}(x)-\phi_{i}(x)\big]^2}{2\Delta_i}
+\left(\frac{\Delta_{i}+\Delta_{i-1}}{2}\right)
\Bigg(-\frac{\big[\phi_{i}(x+1)-\phi_{i}(x)\big]^2}{2\ud x^2}
-\frac{m^2}{2}\phi_{i}^2(x)
-\frac{\lambda}{24}\phi_{i}^4(x)
\Bigg)
\Bigg\},
\end{align}
}where we have adopted a field redefinition as illustrated in \cref{fig:denotion}, and the time differences are denoted as
\begin{align}
\Delta_i=\Big\{
\begin{array}{l}
~~\ud t,~{\rm if}~1\leq i<m; \\
-\ud t,~{\rm if}~m\leq i<2m-1.
\end{array}
\end{align}
In the exponent, there are terms like $\phi^q_1(x)\tilde \phi^{cl}_0(x)-2\phi^q_1(x)\tilde \phi^{cl}_1(x)+\cdots$, where $\tilde \phi^{cl}_0$ and $\tilde \phi^{cl}_1$ can appear.
In fact, an extra $\tilde\phi_2^{cl}(x)\phi^q_1(x)$ term will cancel out these linear-in-$\phi^q_1(x)$ terms, due to the equation of motion (\ref{eq:eom}).
Therefore, we are able to substitute these terms with $\tilde\phi_2^{cl}(x)$ term only, and this simplifies expression (\ref{eq:I}) a lot.
Given that $\phi_1^{cl}$ is part of the specified initial data, we define $\phi_1=\phi^q_1/2$ to ensure that at site 1 only $\phi^q_1$ is included in the dynamical part of the path integral.
To arrive at \cref{eq:I}, we have also used that,
\begin{align}
\phi_{2m-1}=-\phi_1
,\quad
\Delta_{0}=-\ud t.
\end{align}
There are $N_{tot}=N_s(2m-2)$ variables in total, and we will adopt a more compact notation, merging space and time labels into a single integer ${a}$.

For all the field variables $\phi_a$, we start our Monte-Carlo chain for the dynamical part of the path integral from $\tilde{\phi}_a$, the classical critical-point configuration. In subsequent Monte-Carlo steps, these will be changed into new real values $\varphi$. For each such value, the gradient flow equation into the complex plane now reads
\begin{align}
\frac{\ud \phi_{a}}{\ud \tau} = \overline{\frac{\partial {\mathcal I}}{\partial \phi_{a}}},
\end{align}
The Jacobian matrix $J$ itself, defined with element 
$J_{{a}b}=\partial\phi_{a}/\partial \varphi_b$,
evolves along the flow as, 
\begin{align}
\frac{dJ_{{a}b}}{d\tau} = \overline{\frac{\partial^2 {\mathcal I}}{\partial \phi_{a}\partial \phi_l}J_{lb}},
\end{align}
where a summation over index $l$ is understood.

For the Lefschetz Thimble Method, then $J(\tau=0)$ is determined by the eigenvectors of positive eigenvalues \cite{Cristoforetti:2012su} of the Hessian evaluated on the critical point field configuration.
We use the Generalized Thimble Method, then $J(\tau=0)$ is just the identity matrix \cite{Alexandru:2017lqr}.
With these flow equations, one can now apply thimble methods to generate samples for the dynamical part of the path integral.
For more on algorithms based on the Lefschetz Thimble Method, see \cite{Cristoforetti:2012su,Cristoforetti:2013wha,Mukherjee:2013aga}.
And for more on algorithms based on the Generalized Thimble Method, see \cite{Alexandru:2015xva,Alexandru:2015sua,Alexandru:2017oyw,Alexandru:2017czx,Alexandru:2018fqp,Alexandru:2018ngw,Fukuma:2017fjq,Tanizaki:2017yow}.

\begin{figure}[t]
\begin{tabular}{lr}
\hspace{-1.2cm}
\includegraphics[width=0.6\textwidth]{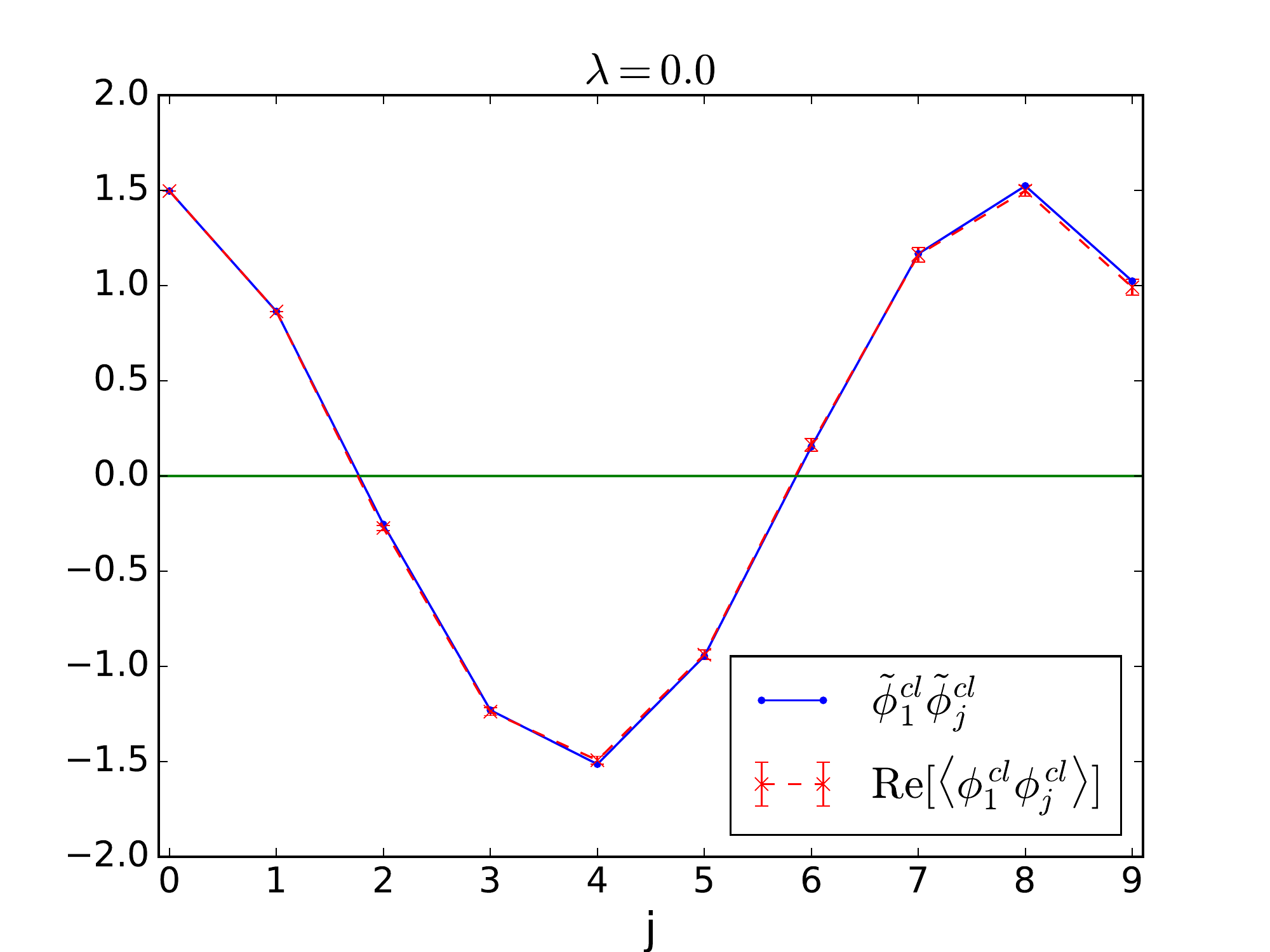} & \hspace{-1.2cm}
\includegraphics[width=0.6\textwidth]{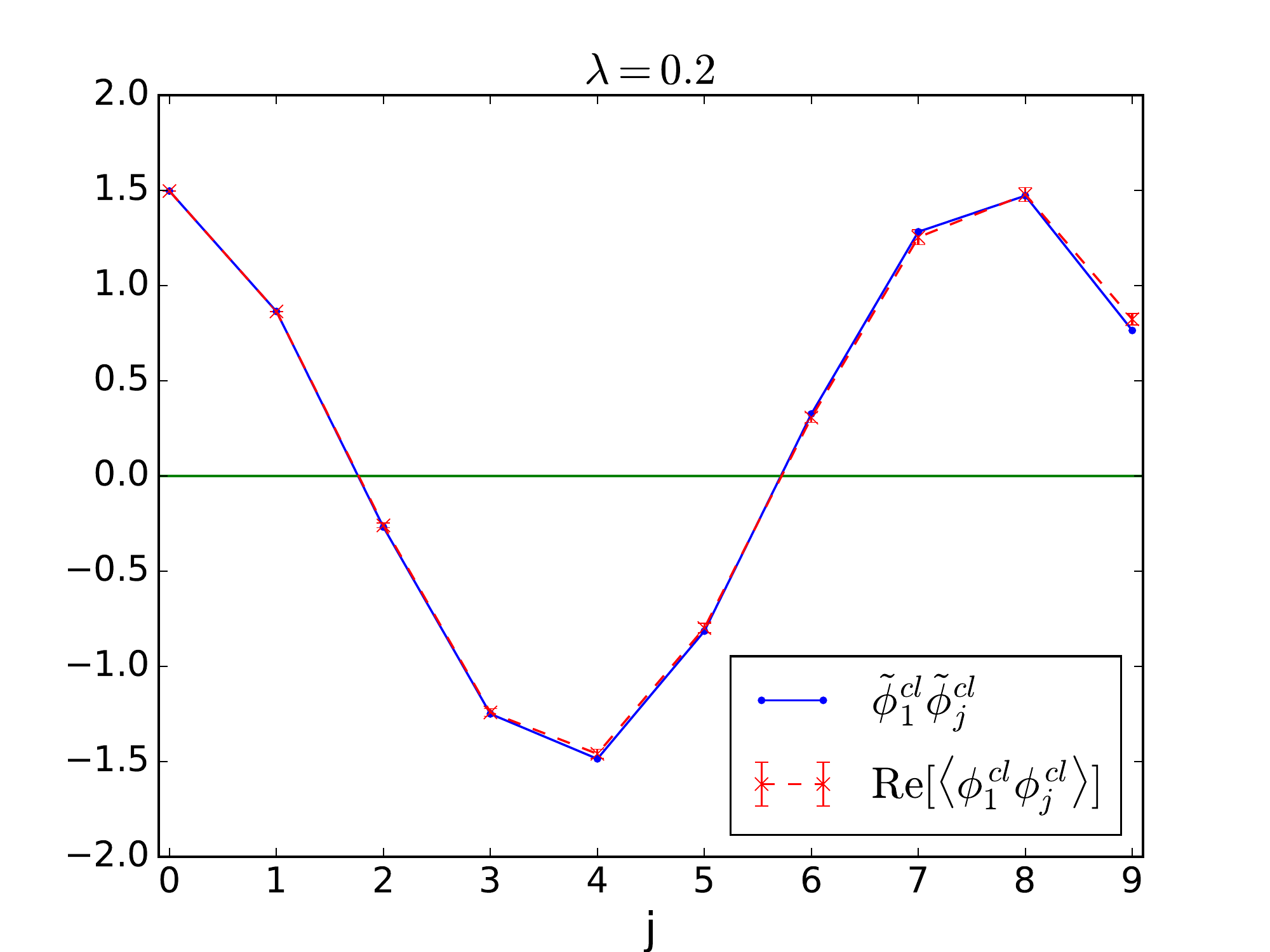}
\end{tabular}
\caption{The classical correlator for a single initial condition, and the corresponding quantum averaged correlator. For $\lambda=0.0$ (left) and $0.2$ (right).}
\label{fig:insideaver}
\end{figure}

Our algorithm can be briefly summarized as follows:\\
\begin{enumerate}
\item Generate an initial value for $\phi_0^{cl}$ and $\phi_1^{cl}$ according to a Gaussian distribution given by \cref{eq:initial}. Determine the critical configuration by solving \cref{eq:eom}.
\item Set $\varphi^{cl}=\phi^{cl}=\tilde\phi^{cl}$, $\varphi^{q}=\phi^{q}=0$ as the starting point of the thimble approach. 
Evolve $\phi$ and $J$ from $\tau=0$ to $\tau=\tau_f$, for some final flow time $\tau_f$
\item To go from the $n$-th to the $n+1$-th configuration in our Monte-Carlo chain for the dynamical part of the path integral, first propose the $(n+1)$-th configuration $\varphi_{n+1}=\varphi_n+\Delta$, where the vector $\Delta$ follows the proposal distribution,
\begin{align}
{\rm Pr}(\varphi_n\to \varphi_{n+1})=\sqrt{\frac{{\rm det }(J_n^\dagger J_n)}{\pi^{N_{tot}}\delta^{2N_{tot}}}}e^{-\Delta^T(J_n^\dagger J_n)\Delta/\delta^2},
\end{align}
with some constant parameter  $\delta$ \footnote{In practice, one can first draw complex vector $\eta$, satisfying Gaussian distribution $\exp(-\eta^\dagger\eta/\delta^2)$, and then $\Delta={\rm Re}(J_n^{-1}\eta)$.}.
\item Use the gradient flow equation to evolve $\phi_{n+1}$ and $J_{n+1}$ from $\tau=0$ to $\tau=\tau_f$.
\item Accept or reject new configuration according the acceptance probability
\begin{align}
&\hspace{5cm}{\rm P}_{acc}(\varphi_n\to \varphi_{n+1})=
 \\
&{\rm min}\{1,\,e^{-{\rm Re}[{\mathcal I_{n+1}}]+ 2\,{\rm ln}\,|{\rm det }\,J_{n+1}|-\Delta^T(J_{n+1}^\dagger J_{n+1})\Delta/\delta^2+{\rm Re}[{\mathcal I_{n}}]- 2\,{\rm ln}\,|{\rm det }\,J_{n}|+\Delta^T(J_{n}^\dagger J_{n})\Delta/\delta^2}\}.
\nonumber
\end{align}
If the new configuration is rejected, choose the $(n+1)$-th configuration to be the same as the $n$-th configuration. 
\item Repeat (3)-(5) until we have enough statistically independent configurations to average over, for this one initial condition realisation.
\item Repeat (1)-(6) for $n_{initial}$ times, to get enough initial conditions to average over (these are statistically independent by construction).
\end{enumerate}

On the thimble approach (3)-(5), we follow the prescription given by \cite{Alexandru:2017lqr}, with the  difference that we perform an LU decomposition for matrix $J$ to calculate its inverse and determinant directly.
Therefore, we can have the acceptance probability with the explicit existence of ${\rm det}\,J$.
With the proposal distribution and acceptance probability above, the obtained samples will admit the probability weight $P=e^{-{\rm Re}[{\mathcal I}]+ {\rm ln}\,|{\rm det }\,J|}$. The numerical effort is substantial, and many technical details, performance tests and detailed numerical investigations will be reported in our future publications.

In \cref{fig:insideaver}, we show the correlator for a single classical trajectory, and compare it to the correlator when averaging over the quantum variables (but without averaging over initial conditions, only step 1-6 of our algorithm). In the left-hand plot for the free theory ($\lambda=0$), in the right-hand plot including interactions ($\lambda=0.2$). We see that the quantum averaging is has only a small effect for the free theory, whereas including a moderate interaction strength there is statistically significant effect, increasing over time.

\subsection{All warmed up: Full quantum evolution}
\label{sec:quantumaverage}

We are now ready to carry out the inner (Monte-Carlo integration on the thimble) and outer (initial conditions) integration together, to find the full quantum correlator, given our initial Gaussian state. The simulations presented here use $n_{initial}=200\sim 60$ initializations, with $(5\sim 20)\times 10^5$ Metropolis updates for single initialisation, in order to give small enough statistical errors. 

\begin{figure}[t]
\begin{tabular}{lr}
\hspace{-1.2cm}
\includegraphics[width=0.6\textwidth]{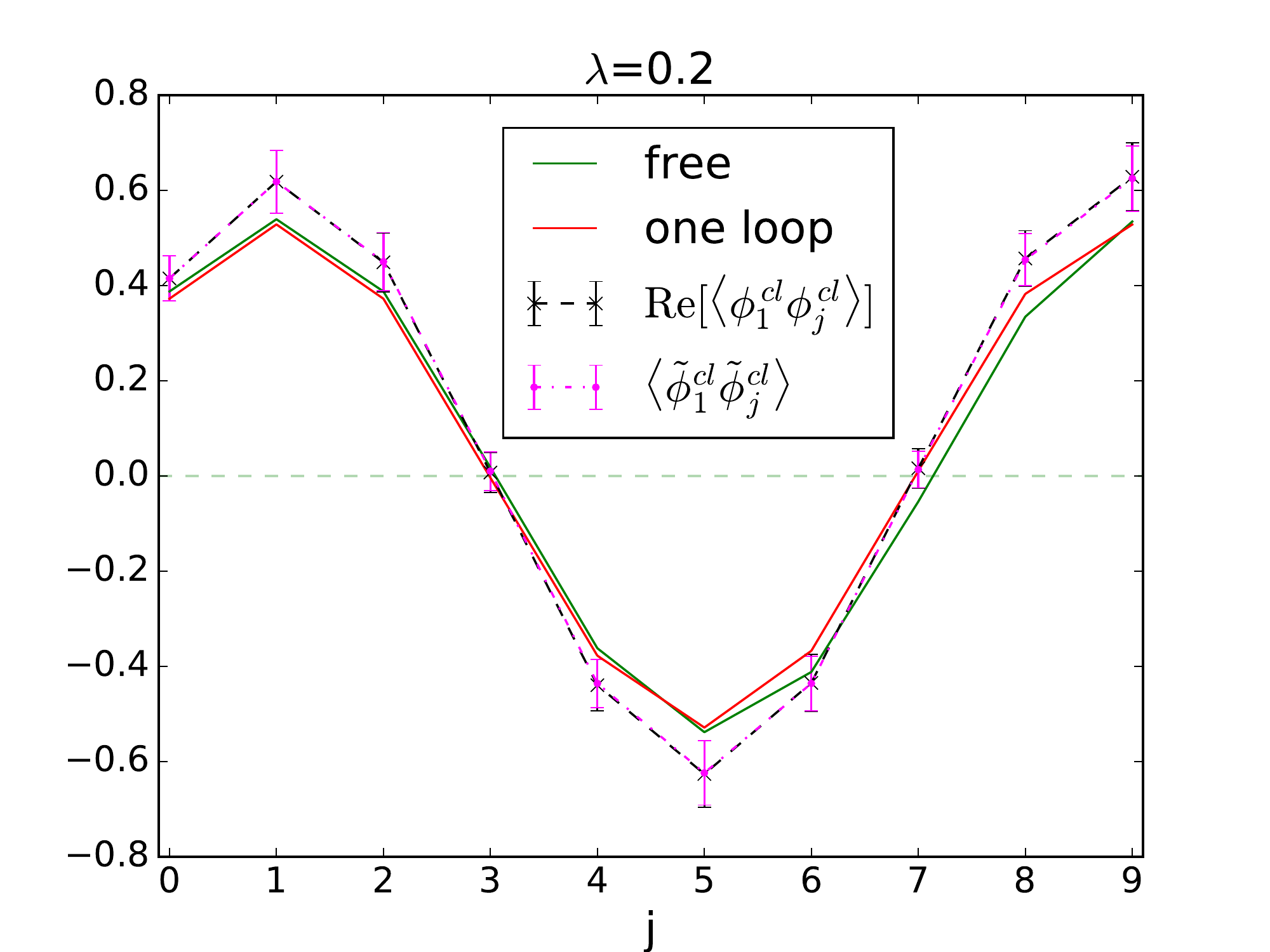} & \hspace{-1.2cm}
\includegraphics[width=0.6\textwidth]{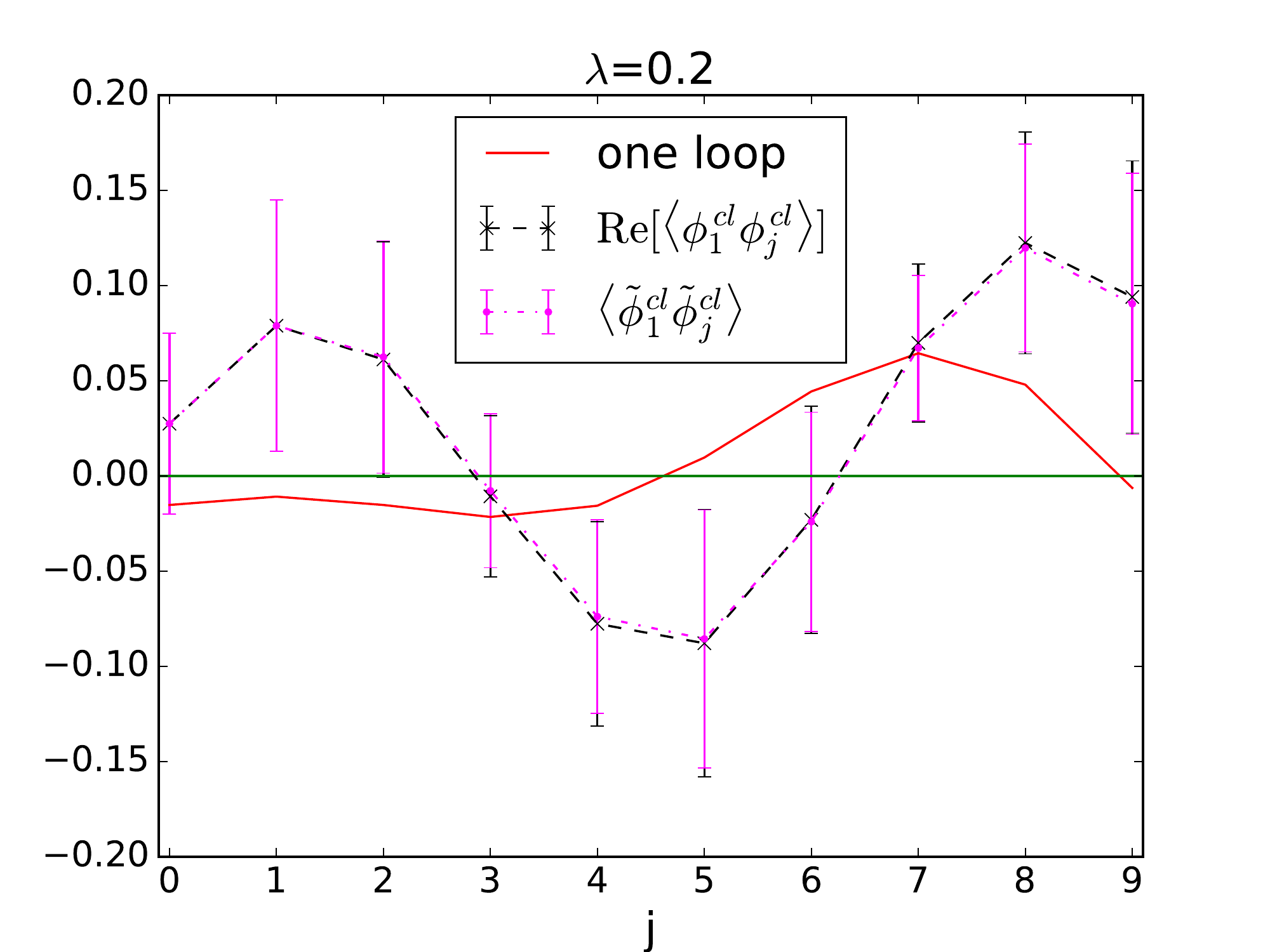}
\end{tabular}
\caption{The full classical-statistical and quantum correlators (cl-cl) for a free and interacting theory at $\lambda=0.2$. The figure on the right shows the result of subtracting the free propagator. The red line is the perturbative 1-loop result. }
\label{fig:fulllambda1}
\end{figure}

\Cref{fig:fulllambda1} (left) shows the two-point cl-cl correlator for the full classical-statistical simulation (pink) and the full quantum simulation (black). Overlaid also the 1-loop perturbative result (in red). \Cref{fig:fulllambda1} (right) arises from subtracting the free propagator, to highlight the contribution from interactions.  We see that the classical-statistical approximation performs very well at these values of the coupling, and that apparently the differences arising from quantum averaging each initial condition (\cref{fig:insideaver}) are in turn largely washed out when averaging over initial conditions. The 1-loop approximation shown in red is distinct from the other two curves, showing that we are not in the extreme small-coupling limit, and so the agreement between classical-statistical and quantum approaches does apply to an interacting system.

We now proceed to increase the coupling $\lambda$, beyond the naively perturbative domain. We show in \cref{fig:fulllambda2} the case $\lambda=4$, where  we can now clearly distinguish the classical-statistical (pink) from the fully quantum result (black). They are both different from the free theory (green) and the 1-loop approximation (red). 
\begin{figure}[t]
\centering
\begin{tabular}{lr}
\hspace{-1.2cm}
\includegraphics[width=0.6\textwidth]{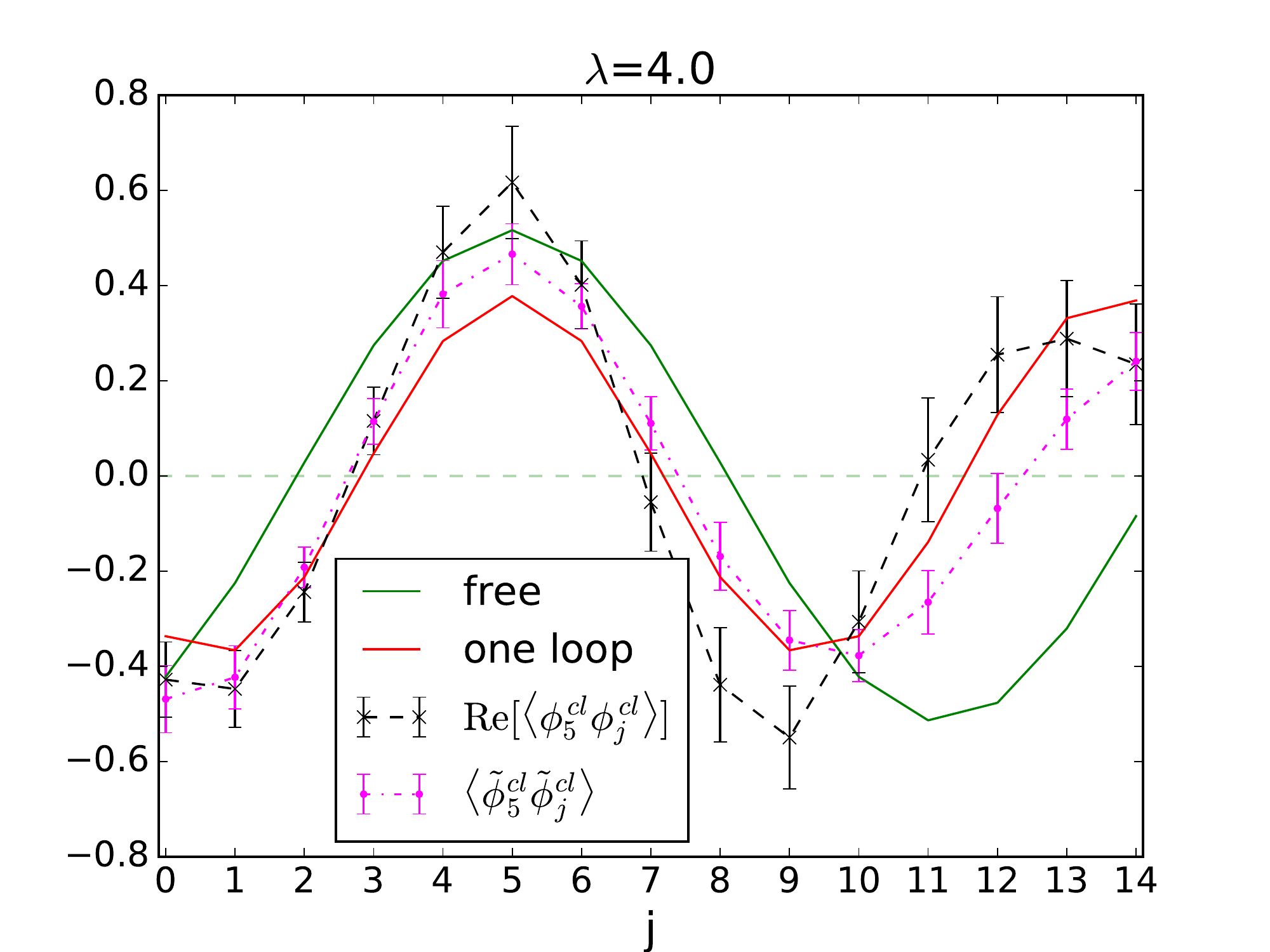} & \hspace{-1.2cm}
\includegraphics[width=0.6\textwidth]{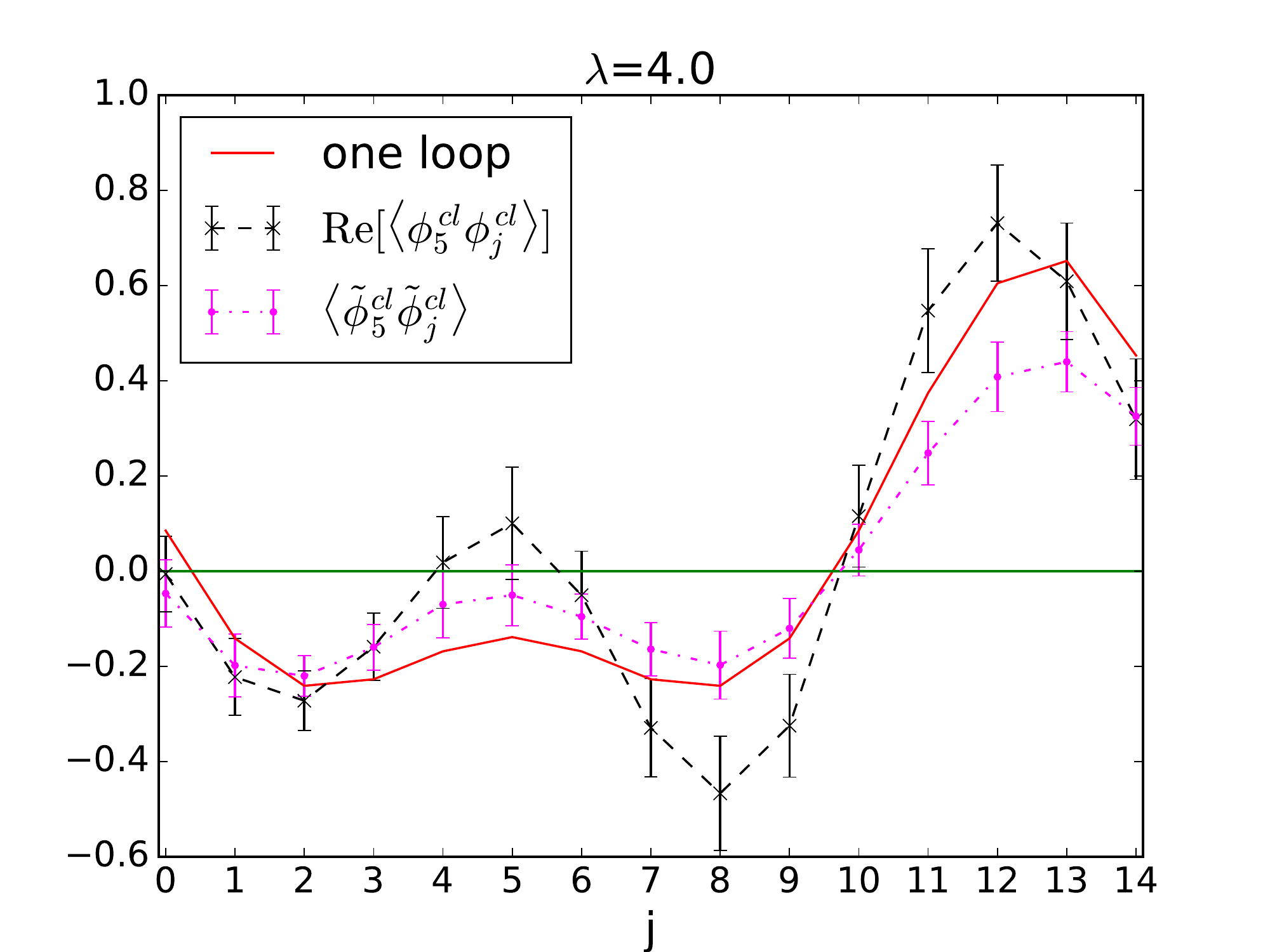}
\end{tabular}
\caption{On the left, the full quantum correlators (cl-cl) for a free and interacting theory at $\lambda=4$. On the right, when subtracting the free propagator.}
\label{fig:fulllambda2}
\end{figure}

\section{Conclusions}
\label{sec:summary}

Real-time quantum dynamics is well-defined in terms of the Schwinger-Keldysh formalism, and although the classical-statistical approximation often does very well in some cases, simulations of truncated Kadanoff-Baym equations have shown that quantum corrections are important in other contexts. 

We have investigated a new, in principle exact, method for computing real-time quantum correlators directly from the path integral. This is possible through Monte-Carlo sampling, as the sign problem inherent to the complex action can be softened by flowing the field variables 
into the complex plane. 

We have presented a number of technical developments necessary to generalise the work of \cite{Alexandru:2016gsd,Alexandru:2017lqr} to initial-value problems. 
For a discrete space-time, we have verified that the scalar field path integral can be separated into two parts: the initial density matrix and the following dynamical part. Under such a separation there exists one and only one critical point, which helps when we implement either the Lefschetz Thimble Method, or the Generalized Thimble Method on the dynamical part.
We use a symmetric discretization of the theory, in both a symmetric Feynman kernel and a symmetric time contour. With such a discretization we can find all the critical points.

To demonstrate the implementation of our approach, we have computed the real-time propagator for a scalar field in 0+1 dimensions, with a Gaussian (free-field) initial condition. We found good statistical convergence, and agreement with the free analytic correlator (up to discretization errors). Once interactions were included and increased we found that we could distinguish from the free case, that the 1-loop perturbative result began to fail, and that for very large couplings, the classical-statistical approximation became unreliable. 

In the present paper we have used the initial density matrix of the free theory, as in this case, we can integrate out $\phi_0^{q}$ explicitly, allowing us to obtain the familiar initial distribution of $\phi_0^{cl}$ and $\dot\phi_0^{cl}$.
There is no difficulty in extending the calculation to the case of a more general density matrix, as long as we know how to generate the initialization for $\phi_0^{cl}$ and $\dot\phi_0^{cl}$.
Note, however, that a density matrix containing $\phi_0^{q}$ and $\phi_1^{cl}$  might still be plagued with the ``sign problem'' owing to the appearance of a factor of $i\phi^{cl}_1\phi^q_0$ in (\ref{eq:phi0qphi1c}). This only affects the density matrix part of the path integral, so the thimble approach may still be used for the remaining dynamical part.
On the other hand, we have also in mind that real physical situations can be modeled by turning on the interaction after the initialization, either instantly or gradually, and the method developed in the present paper can naturally deal with time dependent interaction coefficients.

The computational cost of the thimble approach is $a{\mathcal O}(n^3)$, with $n$ the number of variables and $a$ the number of samples.
By separating the simulation into two parts with $n_1$ and $n_2$ variables respectively, the cost becomes $a_1{\mathcal O}( n_1^3)+ a_1a_2{\mathcal O}( n_2^3)$, corresponding to generating $a_1$ different initializations and for each initialization $a_2$ Monte Carlo samples.
If $a$ is not sensitive to $n$, the cost will be smaller than $a{\mathcal O}((n_1+n_2)^3)$, when $n_1$ and $n_2$ are big numbers.
In fact, if this is the case, we can further separate the path integral into more pieces, with each piece depending only on its predecessor but not successor, as each piece becomes an initial condition for the part that follows it.

We have postponed a number of numerical technicalities, diagnostics of the method and further numerical tests of various aspects of the approach to a future publication. 
Simulations on more general initial conditions and potentials, and in 1+1 dimensions  are also underway.

\vspace{0.4cm}

{\noindent \bf Acknowledgements:}
PMS would like to thank Alexandru for some email correspondence. ZGM would like to thank Prof. Bedaque for useful suggestions on the algorithm. PMS and SW were supported by STFC Grant No. ST/L000393/1 and ST/P000703/1. 
AT and ZGM are supported by a UiS-ToppForsk grant.
The authors were also supported by a ECIU travel grant.
The numerical work was performed on the Abel supercomputing cluster of the Norwegian computing network Notur.

\appendix
\section{Loop corrections}
\label{app:loop}
In this section we shall look at the loop corrections to the two-point functions, and we shall be using the continuum expressions in order to provide approximate expressions to the discrete case. First we look at the loop corrections to the Feynman propagator, and then we will see how the computation is adapted to the $(\phi^{cl},\phi^q)$ basis. 

The Feynman propagator is given in (\ref{eq:feyn_prop}) as $i\hbar\int \frac{\ud\omega}{2\pi}\frac{e^{-i\omega(t_x-t_y)+ip(x-y)}}{\omega^2-\omega_p^2+i\epsilon}$, while the interaction vertex is $-\frac{i\lambda}{4!\hbar}$. The loop correction to the propagator is shown in Fig. \ref{fig:feyn_prop_loop_corr}, where the thick solid lines correspond to the Feynman propagator.
\begin{figure}[h]
\centering
\includegraphics[width=0.3\textwidth]{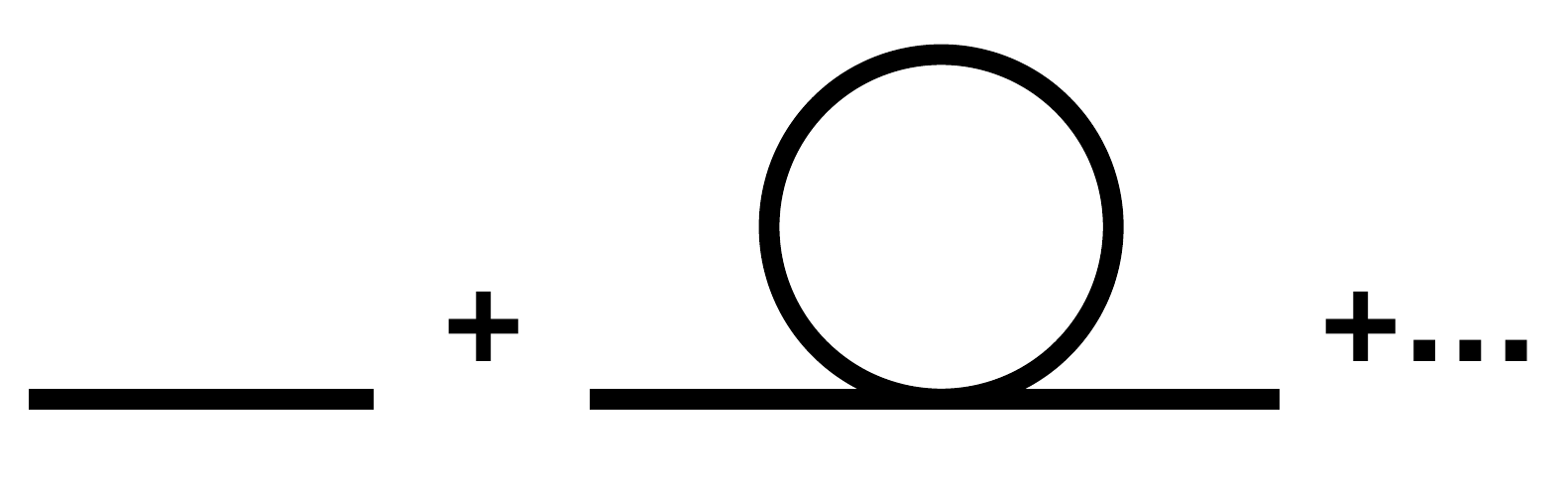}
\caption{Loop correction to the time-ordered two-point correlator, with the thick solid line being the Feynman propagator.}
\label{fig:feyn_prop_loop_corr}
\end{figure}
This may be calculated in zero spatial dimensions as follows.
\begin{align}
\langle T\hat\Phi_1\hat\Phi_2\rangle&=i\hbar\int \frac{\ud\omega}{2\pi}\frac{e^{-i\omega(t_1-t_2)}}{\omega^2-\omega_p^2+i\epsilon}\\\nonumber
	&+12\int\ud t \;i\hbar\int \frac{\ud\omega_1}{2\pi}\frac{e^{-i\omega_1(t_1-t)}}{\omega_1^2-\omega_p^2+i\epsilon}\frac{-i\lambda}{4!\hbar}
					i\hbar \frac{\ud\omega_2}{2\pi}\frac{1}{\omega_2^2-\omega_p^2+i\epsilon}
					i\hbar \frac{\ud\omega_3}{2\pi}\frac{e^{-i\omega_3(t-t_2)}}{\omega_3^2-\omega_p^2+i\epsilon}+...\\\nonumber
	&=i\hbar\int \frac{\ud\omega}{2\pi}\frac{e^{-i\omega(t_1-t_2)}}{\omega^2-\omega_p^2+i\epsilon}\\\nonumber
	&-\frac{\lambda\hbar^2}{2}\int\frac{\ud\omega_1}{2\pi}\frac{\ud\omega_2}{2\pi}e^{-i\omega_1(t_1-t_2)}\frac{1}{\omega_1^2-\omega_p^2+i\epsilon}\frac{1}{\omega_2^2-\omega_p^2+i\epsilon}\frac{1}{\omega_1^2-\omega_p^2+i\epsilon}+...\\\nonumber
	&=i\hbar\int \frac{\ud\omega}{2\pi}\frac{e^{-i\omega(t_1-t_2)}}{\omega^2-\omega_p^2+i\epsilon}\\\nonumber
	&+i\hbar\int\frac{\ud\omega}{2\pi}e^{-i\omega(t_1-t_2)}\frac{1}{\omega^2-\omega_p^2+i\epsilon}\frac{\hbar\lambda}{4\omega_p}\frac{1}{\omega^2-\omega_p^2+i\epsilon}+...\\\nonumber
	&=i\hbar\int \frac{\ud\omega}{2\pi}\frac{e^{-i\omega(t_1-t_2)}}{\omega^2-\omega_p^2-\delta m^2+i\epsilon},
\end{align}
where $\delta m^2=\frac{\hbar\lambda}{4\omega_p}$, and we have used $\int\frac{\ud\omega}{2\pi}\frac{1}{\omega^2-\omega_p^2+i\epsilon}=-\frac{i}{2\omega_p}$.

It is also instructive to use the $(\phi^{cl},\phi^q)$ basis, for which we shall use the continuum expressions to give an approximation to the discrete calculation, and so we start by noting from (\ref{eq:cl-cl}-\ref{eq:q-q}) that the continuum propagators are given by Fig. \ref{fig:prop}, while the interaction vertices are given by Fig. \ref{fig:feyn-int}. 
\begin{figure}[h]
\centering
\includegraphics[width=0.4\textwidth]{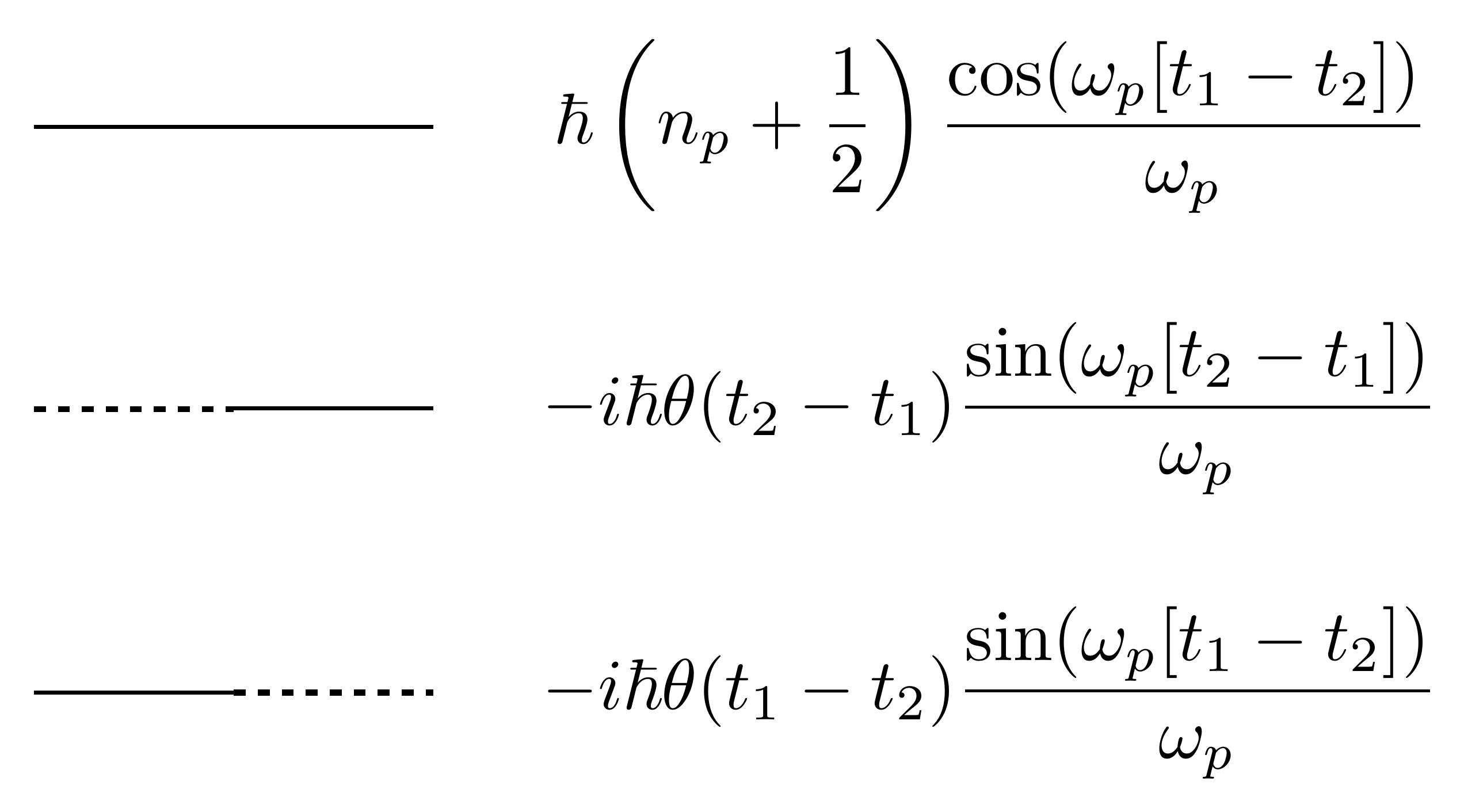}
\caption{Feynman propagators, with the solid line being the $\langle\phi^{cl}\phi^{cl}\rangle_0$ propagator, and the dash-solid line being the $\langle\phi^{q}\phi^{cl}\rangle_0$ propagator.}
\label{fig:prop}
\end{figure}
\begin{figure}[h]
\centering
\includegraphics[width=0.4\textwidth]{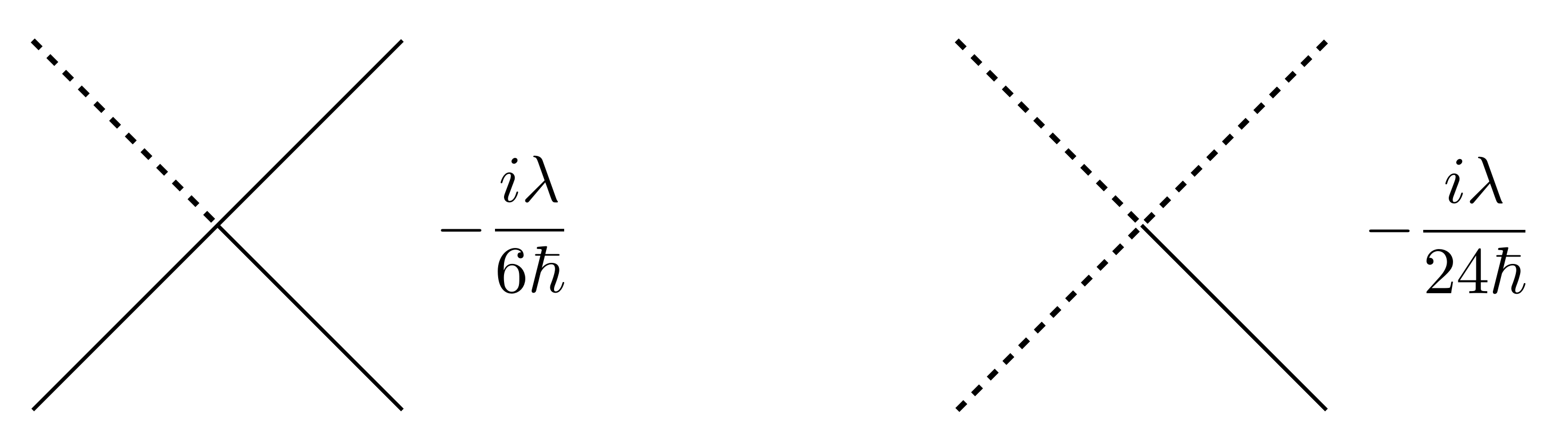}
\caption{Feynman diagrams for the interactions, with the solid line representing $\phi^{cl}$, and the dashed line corresponding $\phi^{q}$.}
\label{fig:feyn-int}
\end{figure}

We now evaluate the loop correction to the advanced propagator, $\langle\phi^{q}\phi^{cl}\rangle$, which we can see in terms of diagrams in Fig. \ref{fig:q_cl_loop}.
\begin{figure}[h]
\centering
\includegraphics[width=0.2\textwidth]{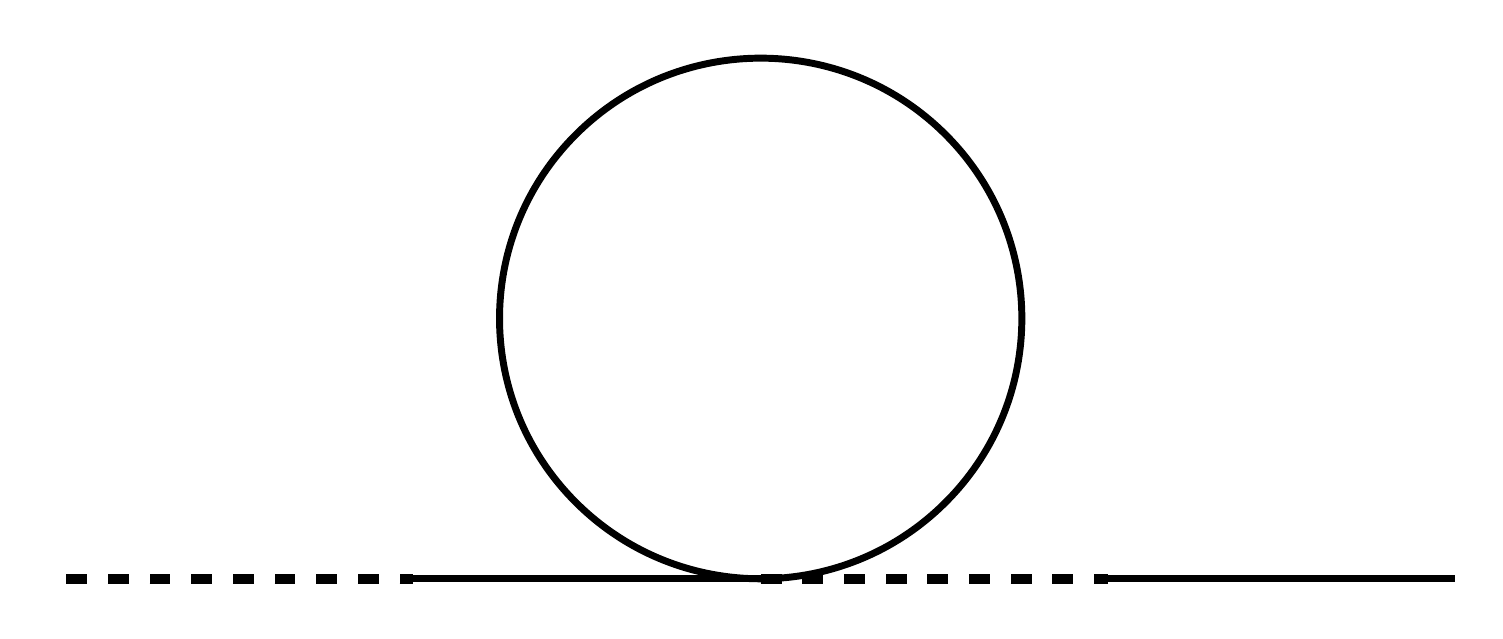}
\caption{Loop correction to the advanced propagator, $\langle\phi^{q}\phi^{cl}\rangle$.}
\label{fig:q_cl_loop}
\end{figure}
\begin{align}
\langle \phi^{q}_1\phi^{cl}_2\rangle&=-i\hbar\theta(t_2-t_1)\frac{\sin(\omega_p(t_2-t_1))}{\omega_p}\\\nonumber
		&+\int\ud t[-i\hbar]\theta(t-t_1)\frac{\sin(\omega_p(t-t_1))}{\omega_p}\frac{\hbar}{2\omega_p}\frac{-i\lambda}{2\hbar}[-i\hbar]\theta(t_2-t)\frac{\sin(\omega_p(t_2-t))}{\omega_p}+...\\\nonumber
		&=-i\hbar\theta(t_2-t_1)\frac{\sin(\omega_p(t_2-t_1))}{\omega_p}\\\nonumber
		&+\int\ud t[-i\hbar]\theta(t-t_1)\frac{\sin(\omega_p(t-t_1))}{\omega_p}\frac{\hbar}{2\omega_p}\frac{-i\lambda}{2\hbar}[-i\hbar]\theta(t_2-t)\frac{\sin(\omega_p(t_2-t))}{\omega_p}+...\\\nonumber
		&=-i\hbar\theta(t_2-t_1)\frac{\sin(\omega_p(t_2-t_1))}{\omega_p}\\\nonumber
		&+i\hbar\theta(t_2-t_1)\frac{\hbar\lambda}{4\omega_p^2}\frac{\sin(\omega_p(t-t_1))-\omega_p(t_2-t_1)\cos(\omega_p(t-t_1))}{\omega_p^2}+...
\end{align}
where we have used the Heaviside theta functions in the propagators  to limit the range of the $t$ integration to $t_1\to t_2$. Now note that the second piece may be written as \mbox{$-i\hbar\theta(t_2-t_1)\frac{\hbar\lambda}{4\omega_p}\frac{\partial}{\partial\omega_p^2}\left[ \frac{\sin(\omega_p(t_2-t_1))}{\omega_p} \right]$}, and so we see that the loop correction corresponds to a correction in $\omega_p^2$ of $\frac{\hbar\lambda}{4\omega_p}$, which is what we found from the Feynman propagator calculation.

The loop correction for the $\langle\phi^q\phi^q\rangle$ correlator is shown, in the generic sense, in Fig. \ref{fig:q_q_loop}, where the blocked out region is any set of lines that follow from the Feynman rules of Figs. \ref{fig:prop} and \ref{fig:feyn-int}. However, what we find in such diagrams is the appearance of a loop of either advanced or retarded propagators, and this vanishes, meaning that there are no perturbative loop corrections to $\langle\phi^q\phi^q\rangle$.
\begin{figure}[h]
\centering
\includegraphics[width=0.3\textwidth]{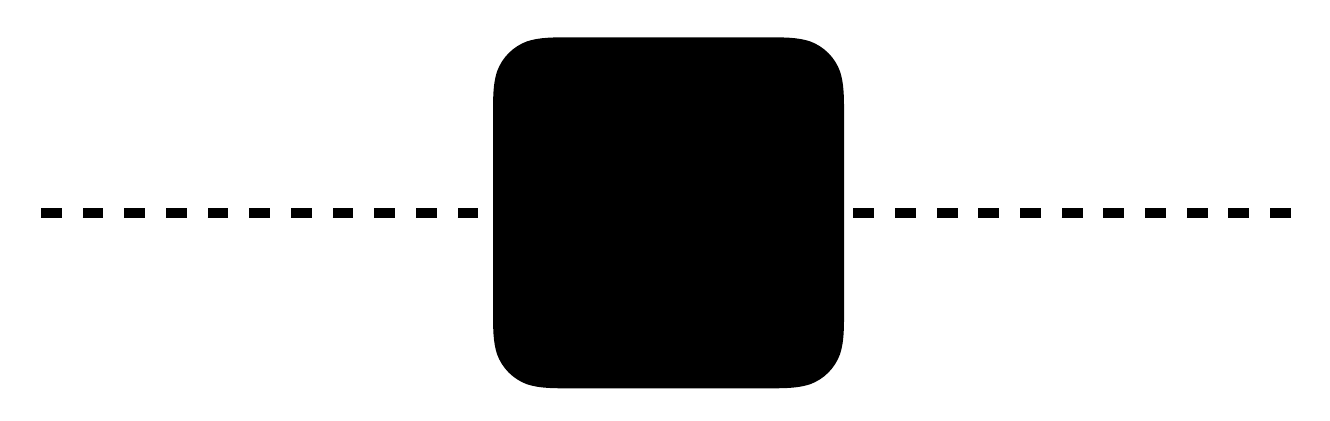}
\caption{There are no non-zero loop corrections to the $\langle\phi^{q}\phi^{q}\rangle$ propagator.}
\label{fig:q_q_loop}
\end{figure}

\end{document}